\documentclass[twocolumn,traditabstract]{aa}  

\usepackage[hyperindex,breaklinks=true, colorlinks, citecolor=blue]{hyperref}  
\usepackage{graphicx}
\usepackage{array}
\usepackage{txfonts}
\usepackage{caption}
\usepackage{lipsum}
\usepackage{hyperref}
\usepackage{url}
\usepackage{natbib}
\usepackage{breqn}
\usepackage{diagbox}
\usepackage{cuted} 
\usepackage{lscape}
\usepackage[mathscr]{euscript}

\newcommand{\correction}[1]{{ \color{red} #1}}
\renewcommand{\correction}[1]{{#1}}

\newcommand{\correctionII}[1]{{ \color{red} #1}}
\renewcommand{\correctionII}[1]{{#1}}

\begin{document}

\title{\correctionII{The} Solar System's passage through the Radcliffe wave during the middle Miocene}
\author{  E. Maconi\inst{1},
          J. Alves\inst{1}, 
          C. Swiggum\inst{1},
          S. Ratzenb\"ock\inst{1,2}, 
          J. Großschedl\inst{1,3,4}, 
          P. K\"ohler\inst{5},
          N. Miret-Roig\inst{1}, 
          S. Meingast\inst{1}, 
          R. Konietzka\inst{6},
          C. Zucker\inst{6}, 
          A. Goodman\inst{6}, 
          M. Lombardi\inst{7},
          G. Knorr\inst{5}, 
          G. Lohmann\inst{5,8},       
          \correction{J. C. Forbes\inst{9},}
          A. Burkert\inst{10},
          \and
          M. Opher\inst{11,12}
          }
\institute{
        University of Vienna, Department of Astrophysics, T\"urkenschanzstraße 17, 1180 Wien, Austria\\ \email{efrem.maconi@univie.ac.at}
        \and
        University of Vienna, Research Network Data Science at Uni Vienna, Kolingasse 14-16, 1090 Vienna, Austria
        \and 
        I. Physikalisches Institut, Universit\"at zu K\"oln, Z\"ulpicher Str. 77, D-50937 K\"oln, Germany
        \and
        \correction{Astronomical Institute of the Czech Academy of Sciences, Boční II 1401, CZ-141 31 Prague, Czech Republic}
        \and
        Alfred-Wegener-Institut Helmholtz-Zentrum f\"ur Polar- und Meeresforschung, 27570 Bremerhaven, Germany
        \and
        Harvard University Dep. of Astronomy and Center for Astrophysics | Harvard \& Smithsonian, Cambridge, MA, USA
        \and
        Università degli Studi di Milano, Dipartimento di Fisica, Via Celoria 16, I-20133 Milano, Italy
        \and
        Department of Environmental Physics and MARUM, University of Bremen, Bremen, Germany
        \and
        \correction{School of Physical and Chemical Sciences, Te Kura Mat\=u, University of Canterbury, Christchurch 8140, New Zealand}
        \and
        University of Munich, Physics Department, Scheinerstrasse 1, D-81679 Muenchen, Germany
        \and
        Radcliffe Institute for Advanced Studies at Harvard University, Cambridge, MA, USA
        \and
        Astronomy Department, Boston University, Boston, MA 02215, USA 
        }

\date{Received 30 August 2024; accepted 8 January 2025}

\titlerunning{\correctionII{The} Solar System's passage through the Radcliffe wave}
\authorrunning{E.~Maconi et al.}

\abstract
    {As the Solar System \correction{orbits} the Milky Way, it encounters various Galactic environments, including dense regions of the interstellar medium (ISM). These encounters can compress the heliosphere, exposing parts of the Solar System to the ISM, \correction{while also increasing the influx of interstellar dust into the Solar System and Earth's atmosphere.} The discovery of new Galactic structures, such as the Radcliffe wave, raises the question of whether the Sun has encountered any of them.}
    {The present study investigates the \correction{potential} passage of the Solar System through the Radcliffe wave \correction{gas structure over the past 30 million years (Myr).}}
    {We used a sample of 56 high-quality, young ($\leq30 \, \mathrm{Myr}$) open clusters associated with a region of interest of the Radcliffe wave to trace its motion back and investigate a potential crossing with the Solar System’s past orbit.}
    {We find that the Solar System’s trajectory intersected the Radcliffe wave in the Orion region. We have constrained the timing of this event to between 18.2 and 11.5 Myr ago, with the closest approach occurring between 14.8 and 12.4 Myr ago. \correction{Notably, this period coincides with the Middle Miocene climate transition \correction{on Earth}, providing an interdisciplinary link with paleoclimatology. \correction{The potential impact of the crossing of the Radcliffe wave on the climate on Earth is estimated.} This crossing could also lead to anomalies in radionuclide abundances, which is an important research topic in the field of geology and nuclear astrophysics.}}
    {} 

\maketitle

\section{Introduction}\label{sec:Intro}

As our Solar System orbits the Milky Way, it encounters different Galactic environments with varying interstellar densities, including hot voids, supernova (SN) blast wave fronts, and cold gas clouds.
The Sun’s passage through a dense region of the interstellar medium (ISM) may impact the Solar System in several ways \citep{Fields2023,Opher2024}. For instance, the enhancement of the ram pressure compresses the heliosphere, exposing some parts of the Solar System to the cold and dense ISM \citep{Miller2022,Miller2024,Opher2024,Opher2024b}. Additionally, the amount of interstellar dust loaded into Earth's atmosphere would increase, potentially enhancing the delivery of radioisotopes \correction{contained in the ISM gas}, such as $^{60}$Fe, via dust grains \citep[see e.g.,][]{Altobelli2005,Breitschwerdt2016,Schulreich2017,Schulreich2023}. This could cause geological radionuclide anomalies \citep{koll2019,Wallner2015,Wallner2021}. Moreover, an increased amount of dust within the Solar System might alter Earth's radiation budget, resulting in a cooling effect  \citep[see e.g.,][]{Shapley1921,Talbot1977,Pavlov2005}. \correction{Therefore, it is crucial to investigate which Galactic environment was encountered by the Sun over its path.}

Understanding the solar neighborhood, with its structures and the physical processes occurring within it, is thus of critical importance. Historically, this understanding has relied on plane-of-the-sky observations - meaning 2D ($l$-$b$) or pseudo-3D projections ($l$-$b$-$v_{\mathrm{R}}$) of the actual underlying 3D structure. However, a new era in astronomy has begun with ESA's \textit{Gaia} mission \citep{Gaia2016}. 
The astrometric data from \textit{Gaia}, complemented with the spectroscopic information on the radial velocity of stars obtained by \textit{Gaia} itself and other previous surveys such as LAMOST, RAVE, GALAH, and SDSS \cite[][]{Luo2015,Majewski2017,Steinmetz2020,Buder2021,Abdurrouf2022}, have opened a new 6D window into the stellar content of the Milky Way and a new understanding of the local ISM. Advanced statistical techniques have enabled the creation of 3D dust maps, extending up to several kiloparsecs from the Sun and achieving parsec-scale resolution for the nearby environment \citep[see e.g.,][]{Green2019,Leike2020,Lallement2019,Lallement2022,Vergely2022,Edenhofer2024}. The analysis of \textit{Gaia}-era 3D dust maps and \correction{molecular clouds' catalogs solely based on 3D positional data \citep[see e.g.,][]{Zucker2019,Chen2020,Dharmawardena2023,Cahlon2024}} has unveiled structures such as the Radcliffe wave \citep{Alves2020} and the Split \citep{Lallement2019}, which were previously thought to constitute \correction{a ring-like structure around the Sun, named} the Gould Belt \citep{Gould1874}, due to misleading projection effects. 

In this paper, we investigate the possible encounter between the Sun and the Radcliffe wave. The Radcliffe wave \citep[][]{Alves2020} is a narrow (aspect ratio of 1:20) and coherent $\sim$2.7-kpc-long sinusoidal gas structure, which comprises many known star-forming cloud complexes, such as CMa, Orion, Taurus, Perseus, Cepheus, North America nebula, and Cygnus. This gas structure, with an estimated mass of $3 \cdot 10^6\,\mathrm{M}_\odot$, appears to coherently oscillate like a traveling wave \citep{Konietzka2024} and it is thought to be part of the Galaxy spiral structure \citep{Swiggum2022}. \correction{We used} recent open cluster catalogs to identify a subset of young ($\leq30\,\mathrm{Myr}$), open clusters associated with the Radcliffe wave. \correction{By leveraging the new information regarding the 3D structure of the local ISM and the 3D spatial motions of the selected open clusters, used as tracers of the motion of the primordial clouds out of which they were born, we investigate potential interactions of our Solar System and the Radcliffe wave. Additionally, we discuss the possible geological signatures and climate effects that such an interaction could produce.}


This paper is organized as follows. In \correction{Sect.}~\ref{sec:Data}, we outline the data used for the study. In \correction{Sect.}~\ref{sec:Methods}, we describe the selection of the Radcliffe wave clusters, the estimation of their properties, such as age and mass, along with the properties of their parental clouds, and the method for orbit integration.
We discuss the results and their interdisciplinary connections with other fields of study in \correction{Sect.}~\ref{sec:Results_Discussion} and we summarize our findings in \correction{Sect.}~\ref{sec:Conclusions}.

\section{Data}\label{sec:Data}

The Galactic cartesian coordinates ($X,\,Y,\,Z$) of the molecular clouds constituting the Radcliffe wave structure are taken from the studies by \cite{Zucker2019} and \cite{Alves2020}. The molecular clouds catalog by \cite{Zucker2019} was constructed using a Bayesian statistical method, incorporating optical and near-infrared photometry, along with astrometric data, from the Pan-STARRS survey \citep{Chambers2016,Flewelling2020}, the 2MASS survey \citep{Skrutskie2006}, the NOAO source catalog \citep{Nidever2018}, and parallaxes from the second \textit{Gaia} Data Release \citep{Arenou2018,Gaia2018}. \correction{We used the positions of the clouds that constitute the Radcliffe wave to identify young clusters that may be associated with it.}

For the clusters, we primarily used the recent open star cluster catalog by \cite{Hunt2023} \correction{(hereafter, HR23)}. This catalog was constructed by applying the Hierarchical Density-Based Spatial Clustering of Applications with Noise routine \citep{McInnes2017} (HDBSCAN) on the \textit{Gaia} DR3 astrometrical data \citep{Gaia2023} and the results were validated through a statistical density test and a Bayesian convolutional neural network. The catalog comprises a total of 7166 star clusters and provides a broad array of parameters. Of particular importance for our research are the sky positions, parallaxes, proper motions, radial velocities (RVs), estimated ages, extinctions, and stellar membership lists. To get better statistics for the full 6D phase space, we added additional RV data to the existing \textit{Gaia} DR3 RVs. We cross-matched the stellar members of each cluster to the following surveys: APOGEE-2 SDSS DR17 \citep{Abdurrouf2022}, GALAH DR3 \citep{Buder2021}, RAVE DR6 \citep{Steinmetz2020}, \textit{Gaia} ESO DR6 \citep{Randich2022}, LAMOST DR5 \citep{Zhao2012,Tsantaki2022}, and two RV compilations \citep{Gontcharov2006,Torres2006}. In cases where a star is present in multiple surveys, we selected the \correction{RV} value with the lowest uncertainty. Before recomputing the clusters’ median RVs, we first excluded stars with RV errors greater than $5 \, \mathrm{km}\,\mathrm{s}^{-1}$ and, secondly, we applied sigma-clipping using 3-sigma around the cluster median. To further ensure the accuracy of our results, we imposed a minimum requirement of at least five stars used in the computation of each cluster's median RV and we considered only those clusters with \correction{RV} errors below $3 \, \mathrm{km}\,\mathrm{s}^{-1}$. Clusters that have only three or four stellar members with RVs were included again \correction{if the standard error is below} $1.2 \, \mathrm{km}\,\mathrm{s}^{-1}$, since these show a very narrow distribution in RV space. \correction{We then computed each cluster's mean Galactic cartesian velocities ($U,\,V,\,W$) and the corresponding standard error (see Table~\ref{tab:RWclusters-6Dinfo-allClusters}).} To ensure the positional accuracy of our results, we considered only those clusters with positional errors below 50 pc for the $XYZ$ positions. Subsequently, we checked that none of the discarded clusters held relevance for our findings. Moreover, we compared our selection to the one used \correction{in a recent paper that investigated the Radcliffe wave's motions} \citep{Konietzka2024}. \correction{In that study, the authors used} a compilation of several cluster catalogs \citep{Sim2019,Liu2019,Cantat-Gaudin2020,Szilagyi2021,Hao2022,He2022} \correction{that predated HR23.} We found that seven clusters of relevance to this study were missing from our selection. \correction{We computed the median positions and velocities of these additional clusters using the membership lists from the auxiliary cluster catalogs and cross-matches with the mentioned RV catalogs, deploying the same procedure as outlined above.} As a result, four additional clusters passed our quality check: \mbox{CWNU 1028}, NGC 1977, OC 0340, and UBC 207. \correction{The Galactic positions and velocities of the clusters, reported in Table~\ref{tab:RWclusters-6Dinfo-allClusters}, were then used to \correction{integrate} their orbits in the past. More information are provided in Sect.~\ref{sec:Methods}.}

\begin{figure*}
    \centering
    \includegraphics[width=1.0\textwidth]{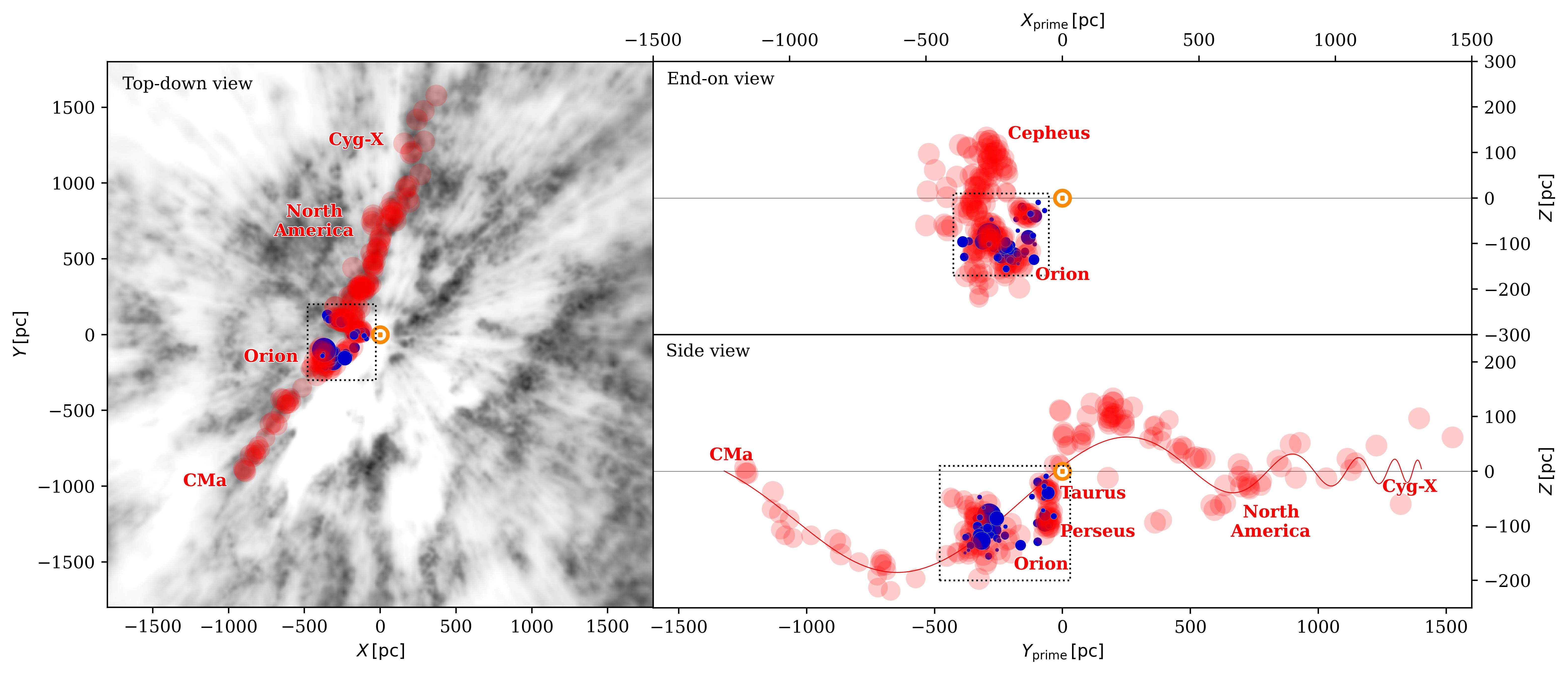}
    \caption{Overview of the Radcliffe wave and selected clusters, \correction{in a heliocentric Galactic cartesian frame}. The Sun is placed at the center and its position is marked with a golden-yellow $\odot$. The red dots denote the molecular clouds and tenuous gas bridge connections that constitute the Radcliffe wave \citep{Zucker2019,Alves2020}. The blue points represent the 56 open clusters associated with the region of the Radcliffe wave that is relevant for this study. The size of the circles is proportional to the number of stars in the clusters. In the $XY$-plane on the left panel, the gray scale represents a dust map \citep{Vergely2022} of the solar neighborhood integrated over the $z$-axis. To properly visualize the Radcliffe wave's structure, the $XY$-plane has been rotated counterclockwise by 120$^\circ$ for the $ZX_{\mathrm{prime}}$-plane (top-right panel) and by 30$^\circ$ for the $ZY_{\mathrm{prime}}$-plane (bottom-right panel), for an observer facing the Galactic anticenter. In the $ZY_{\mathrm{prime}}$-plane, the red-solid line represents the Radcliffe wave's best fit model \citep{Konietzka2024}.}
    \label{fig:RWoverview}
\end{figure*}

\section{Methods}\label{sec:Methods}

\subsection{Identification of the Radcliffe wave’s clusters}\label{sec:RW_clusters_identification}

\correction{We identified the clusters that could be associated with the Radcliffe wave by using the catalog of molecular clouds that constitute this structure \citep[][]{Alves2020} and the cluster sample described in Sect.~\ref{sec:Data}.}
As a first step, we considered the clusters that, at the present day, are within 60 pc to the major molecular clouds and tenuous gas connections comprising the Radcliffe wave. \correction{We opted for this conservative threshold, as the radius of the Radcliffe wave has been estimated to be around 50 pc by \cite{Konietzka2024}. Extending the threshold to 60 pc accounts for positional uncertainties that might bring clusters into this range. We chose not to include clusters farther away as their connection to the Radcliffe wave would be more uncertain. Additionally, we investigated the clusters that are located between 60 and \mbox{100 pc}, and we verified that they are not relevant to our results.} Using this \correction{distance} criterion, we initially selected 104 clusters. 
As a second step, since we are interested in relatively young clusters whose motion may still be related to the gas phase of the Radcliffe wave, we applied an age cut and only considered those with an age estimate smaller than 30 Myr. We computed the ages of the clusters using the \textit{Gaia} DR3 photometry, employing the procedure described in Sect.~\ref{sec:age_mass_computation} \correction{and Appendix~\ref{app:age_computation}}. After applying the age cut, we were left with a total of 74 clusters that we consider to be associated with the Radcliffe wave. 
\correction{As a final selection step, we did a preliminary orbit integration over the past 30 Myr (for the integration details, see Sect.~\ref{sec:orbits_traceback}) to check which clusters come closer than 300 pc to the Sun in this time frame. This left us with a sample of 56 clusters.}

In Fig.~\ref{fig:RWoverview}, we show the present-day positions of the 56 identified clusters, along with the molecular clouds comprising the Radcliffe wave. The heliocentric positions and velocities of the clusters are listed in Table~\ref{tab:RWclusters-6Dinfo-allClusters}. 
\correctionII{The selected clusters belong to the Taurus, Perseus, and Orion star-forming regions of the Radcliffe wave (see Table~\ref{tab:RWclusters-properties}).}
\correction{The cluster names used in this work are listed in Table~\ref{tab:RWclusters-properties}. We primarily use the names from the assembled cluster catalogs, while} we have renamed some of the selected clusters with their previously defined and more widely recognized names, in cases where a significant cross-match of their stellar members aligns with those of previous studies \citep{Chen2020b,Pavlidou2021,Krolikowski2021}.

\correction{We used the 3D positions and velocities of the selected clusters to estimate their past orbits in the Milky Way as well as the previous orbits of the molecular clouds out of which they formed.
The past positions of the gas clouds were inferred from the pre-birth trajectories of the clusters associated with them. This is based on the momentum conservation principle, which implies that the velocities of newly formed and young clusters are correlated with those of the center of mass of the natal gas \citep[see e.g.,][]{Fernandez2008,Tobin2009,Hacar2016,Grossschedl2021,Konietzka2024}. 
As the trajectories of these clusters appear to be interconnected rather than independent, exhibiting common motion (see also Sect.~\ref{sec:ResDisc_crossing}), and as they are all part of the Radcliffe wave, we interpret the orbits of their natal clouds as tracers of a larger gas complex. In this view, the gas clouds should not be considered as isolated objects but rather as parcels of a bigger gas structure. The fragmentation of this structure produced several bound, dense gas clumps where the analyzed star clusters subsequently formed. For simplicity, we adopt this decomposition into distinct ``clouds'' and refer to these parcels of gas as clouds throughout the paper. Notable examples within the analyzed region include the \mbox{Orion A} and \mbox{Orion B} giant molecular clouds, which extend for about 200 pc, each containing many embedded clusters. 
We acknowledge that this modeling approach simplifies the real structure and evolution of the gas clouds, but it is the only one applicable in this case since the exact shape of a cloud in the past cannot be recovered, as the gas is not rigid and its distribution is influenced by multiple mechanisms (see also Sect.~\ref{sec:gas_cloud_properties}). Considering these systematic uncertainties, the trajectories of the clusters and their parental clouds should be considered as preliminary estimates of the actual cloud paths.
We refer the reader to Sects.~\ref{sec:age_mass_computation},  \ref{sec:gas_cloud_properties}, and \ref{sec:orbits_traceback}, for more details on the properties of the clusters, the properties of the parental clouds, and the orbit integration, respectively.
}


\subsection{Estimation of ages and masses of the Radcliffe wave’s clusters} \label{sec:age_mass_computation}

In this paper, we mainly used the HR23 cluster catalog, which already provides a homogeneous estimate of cluster ages and extinctions, using an approximate Bayesian neural network model which was trained on simulated data. However, as the age estimation was not the main focus of that paper, we recomputed the ages with a more robust and slightly more computationally expensive method.
This method uses the recently developed {\tt Python} package {\tt Chronos} \citep[see][for more details]{Ratzenboech2023b}, which is capable of estimating the age, extinction, and metallicity of a cluster by performing a Bayesian \correction{fit of each cluster’s stellar members to theoretical model isochrones. We decided to use the PARSEC models \citep{Bressan2012,Nguyen2022} in combination with \textit{Gaia} DR3 photometry. We assumed the clusters to have the same metallicity, as they are young ($\leq 30 \, \mathrm{Myr}$) and spatially close. We adopted a solar metallicity, since the chemical composition of Radcliffe wave clusters has been shown to be compatible with that of the Sun \citep[see e.g.,][]{Alonso2024}.}
We also recomputed the ages for \correction{the clusters not present in HR23 (see Sect.~\ref{sec:Data} for more details).}
In Table~\ref{tab:RWclusters-properties}, we report the estimated ages for the 56 clusters of the Radcliffe wave that we are considering in this work. In Fig.~\href{https://doi.org/10.5281/zenodo.14626660}{A.1} and \href{https://doi.org/10.5281/zenodo.14626660}{A.2}, we show the color-magnitude diagrams \correction{that are used for isochrone fitting.} In Fig.~\href{https://doi.org/10.5281/zenodo.14626660}{A.3}, we compare the ages of the clusters computed in this work with the ones provided by the source catalogs. We refer the reader to Appendix~\ref{app:age_computation} for more details. 

\correction{We estimated the mass of a given cluster by summing the masses of its stellar members, which were determined from the isochrone fitted by {\tt Chronos}.}
To correct for the incompleteness of the cluster's members, possibly due to observational limits or stellar evolution, \correction{we compared the measured mass distributions to the initial mass function (IMF) by \citet{Kroupa2001}.
By minimizing the total mass difference between a given IMF and the inferred one, we obtained the IMF that best fits our data.} This minimization was performed within the mass range of 0.3 and $2 \, \mathrm{M}_{\odot}$, which is determined by the completeness limits of \textit{Gaia} data \citep{Meingast2021,Gaia2023}. 
The low-mass end limit of the Kroupa IMF was set to $0.03 \, \mathrm{M}_{\odot}$, to account for objects that are below the hydrogen-burning limit. 
We did not specify a high-mass bound to account for potentially missing massive sources that are either too bright for \textit{Gaia} or have undergone SN explosion. 
In Table~\ref{tab:RWclusters-properties}, we list the total masses of the clusters obtained by summing the masses of their members and those derived from the best-fit mass functions. In Fig.~\href{https://doi.org/10.5281/zenodo.14626660}{A.4}, we show the best-fit mass functions, together with the \correction{inferred} ones, for each of the clusters under analysis.

\correction{The reported cluster masses should be regarded as lower limits. This is due to various uncertainties, including the limitations of the HDBSCAN clustering algorithm. These limitations also subsequently affect the cloud radii and estimated number of SN (see Sect.~\ref{sec:gas_cloud_properties} and Appendix~\ref{app:SN_computation}).}
The HDBSCAN algorithm indeed may not select all stellar members of each cluster, even if a star falls within a mass range where \textit{Gaia} data is considered complete. This issue is inherent to all clustering algorithms, as they operate based on their own assumptions that may not fully encompass the wide range of cluster shapes, densities, and sizes. The presence of this systematic bias becomes evident when comparing different solutions from various clustering methods and attempts within the same region \citep[see][for details]{Ratzenboech2023a}.

\subsection{Properties of the parental clouds associated with the clusters}\label{sec:gas_cloud_properties}

\correction{For the times preceding the birth of a given cluster during the orbital tracebacks, we assumed the primordial clouds of the Radcliffe wave to have an onion-like structure. Each cloud was modeled as a set of concentric spheres with radii of 20, 30, 40, and \mbox{50 pc}, representing the denser central parts and lower-density outskirts of each cloud.}
This modeling approach allows us to simplify the complex and varied structures exhibited by gas clouds. Indeed, molecular gas is typically organized into distinct, \correction{filamentary} clouds \citep[see e.g.,][]{Andre2010,Molinari2010,Li2013,Zucker2018,Imara2023} arranged hierarchically, ranging from giant complexes spanning up to 100 pc, down to smaller, denser cores of few parsec in size \citep[see e.g.,][]{Ferriere2001,Motte2018}. 
These clouds can be surrounded by more diffuse gas \citep[see e.g.,][]{Snow2006} and are constantly influenced by various physical mechanisms acting on different scales. These mechanisms include Galactic mechanisms \citep[see e.g.,][]{Inutsuka2015}, stellar feedback \citep[see e.g.,][]{Walch2015,Grossschedl2021,Posch2023}, and magnetic fields \citep[see e.g.,][]{Hennebelle2019}. \correction{Our approach is thus motivated by the fact that we cannot determine the past shapes of the clouds from the data.}

In addition to the onion structure modeling, which aims to include the extension of both the tenuous gas and the central parts of the parental clouds associated with the clusters, we computed an estimate of the mass and radius of the densest part \correction{($n_{\mathrm{H}} \geq 40\,\mathrm{cm}^{-3}$)} of the gas clouds. 
We estimated the mass of the gas cloud associated with a given cluster ($M_{\mathrm{cloud}}$) as

\begin{equation}
    M_{\mathrm{cloud}} = \frac{M_{\mathrm{stars}}}{\mathrm{SFE}} \, ,
\end{equation}
where $M_{\mathrm{stars}}$ is the stellar mass of the cluster corrected for incompleteness, as described in the Sect.~\ref{sec:age_mass_computation}, and SFE is the star formation efficiency for the Radcliffe wave, which is assumed to vary between 1\% and 3\% \citep[see e.g.,][]{Kennicutt2012,Swiggum2022}, \correction{though higher values have also been reported for the SFE in molecular clouds \citep[see e.g.,][]{Chevance2020}.}

\correction{We used a mass-size relation to estimate the radius of an equivalent sphere with the same mass as our estimated cloud masses. Such relations have been studied both observationally and numerically, starting with the work of R.B. Larson \citep{Larson1981}, which was recently updated with the help of \textit{Gaia} data, delivering a new 3D perspective for molecular clouds (see also Sect.~\ref{sec:Intro}). It has been found \citep{Cahlon2024} that the masses of the clouds scale in relation to their volume ($M_{\mathrm{cloud}} \propto r_{\mathrm{cloud}}^3$) when 3D data are used, whereas they scale with the area ($M_{\mathrm{cloud}} \propto r_{\mathrm{cloud}}^2$) when the 3D data are projected onto a 2D plane. This is consistent with the predictions of previous theoretical and numerical studies \citep[see e.g.,][]{Shetty2010,Beaumont2012,Ballesteros-Paredes2019}. Therefore,} we estimated the radius of the densest part for the parental clouds using the observationally based 3D mass-size relation from \cite{Cahlon2024},

\begin{equation}
    M_{\mathrm{cloud}}(r) = 7 \, \mathrm{M}_{\odot} \left( \frac{r}{\mathrm{pc}} \right)^{2.9} \,.
\end{equation}
We list the resulting cloud radii and masses in Table~\ref{tab:RWclusters-properties}. By assuming a SFE of 1\% (3\%), the estimated masses of the clouds span from $1.44 \times 10^5 \, \mathrm{M}_{\odot}$ ($0.48 \times 10^5 \, \mathrm{M}_{\odot}$) for lambda-Ori to $750 \, \mathrm{M}_{\odot}$ ($250 \, \mathrm{M}_{\odot}$) for L1546. The corresponding estimated radii for these clouds are 30.7 pc (21.0 pc) and 5.0 pc (3.4 pc), respectively. As outlined in Sect.~\ref{sec:age_mass_computation}, \correction{the cloud radii and masses are likely underestimated and should be considered as lower limits, resulting from the incompleteness of stellar members in the cluster catalogs.}

\subsection{Orbit integration} \label{sec:orbits_traceback}

We estimated the past orbits of the clusters, \correction{the associated clouds}, and the Sun using the Galactic dynamics package {\tt galpy} \citep{Bovy2015} in combination with the {\tt Astropy} package \citep{astropy}. {\tt galpy} offers the possibility to numerically integrate orbits over different Milky Way potentials and initial conditions, such as the  Galactocentric distance, the Sun's height above the disk mid-plane, and the Sun's velocity. 

For our study, we used {\tt galpy}’s {\tt MWPotential2014} as a model for the Milky Way’s gravitational potential. This model includes a bulge, a disk, and a halo component that are modeled as a power-law density profile with an exponential cut-off, a Miyamoto-Nagai potential, and a Navarro–Frenk–White profile, respectively \citep[see][for details]{Bovy2015}. We assumed a solar Galactocentric radius of $R_{\odot} = 8.33 \, \mathrm{kpc}$ \citep{Gillessen2009} and a vertical position of $z_{\odot}= 27 \, \mathrm{pc}$ \citep{Chen2001}. The Sun’s velocity relative to the Local Standard of Rest, whose circular velocity is set to the default {\tt galpy} value of $220 \, \mathrm{km}\,\mathrm{s}^{-1}$, is  ($U_\odot,\, V_\odot,\, W_\odot$) = (11.1, 12.24, 7.25)$\,\mathrm{km}\,\mathrm{s}^{-1}$ \citep{Schoenrich2010}. These parameters are internally used by {\tt galpy} to change the reference frame \correction{from the Sun's coordinate system to the Galactic center's coordinate system}  \correction{before performing} the orbit integration. The initial positions and velocities of the clusters, defined using the {\tt Astropy} package, serve as input for {\tt galpy}’s orbit module. Our integration covered the past 30 Myr, with a time-step of 0.03 Myr. We employed {\tt galpy}’s {\tt dop853-c} method, a Dormand-Prince integrator known for its reliability and speed. 

We addressed the \correction{statistical} uncertainties in the positions and velocities of the Sun and clusters by integrating their orbits 1000 times, each using new sets of data obtained by sampling the uncertainty distribution with a Monte Carlo sampling method. The initial positions and velocities of the considered clusters with respect to the Sun, along with the errors, are listed in Table~\ref{tab:RWclusters-6Dinfo-allClusters}, while the errors on the solar parameters are sourced from the provided references.
Being aware of the fact that there is not a unique definition of the solar parameters and of the Milky Way potential, in Appendix~\ref{app:initialCondTest} we tested the effect that different initial conditions have on our results. We conclude that the past trajectories of the Sun and the clusters, and their relative distances, do not significantly vary over the last 30 Myr when altering the described parameters. This can be attributed to the relatively short integration time considered, as found from prior studies \citep[see e.g.,][]{Miret-Roig2020}, and supports the robustness of our conclusions.

\begin{figure}
    \centering
    \includegraphics[width=0.495\textwidth]{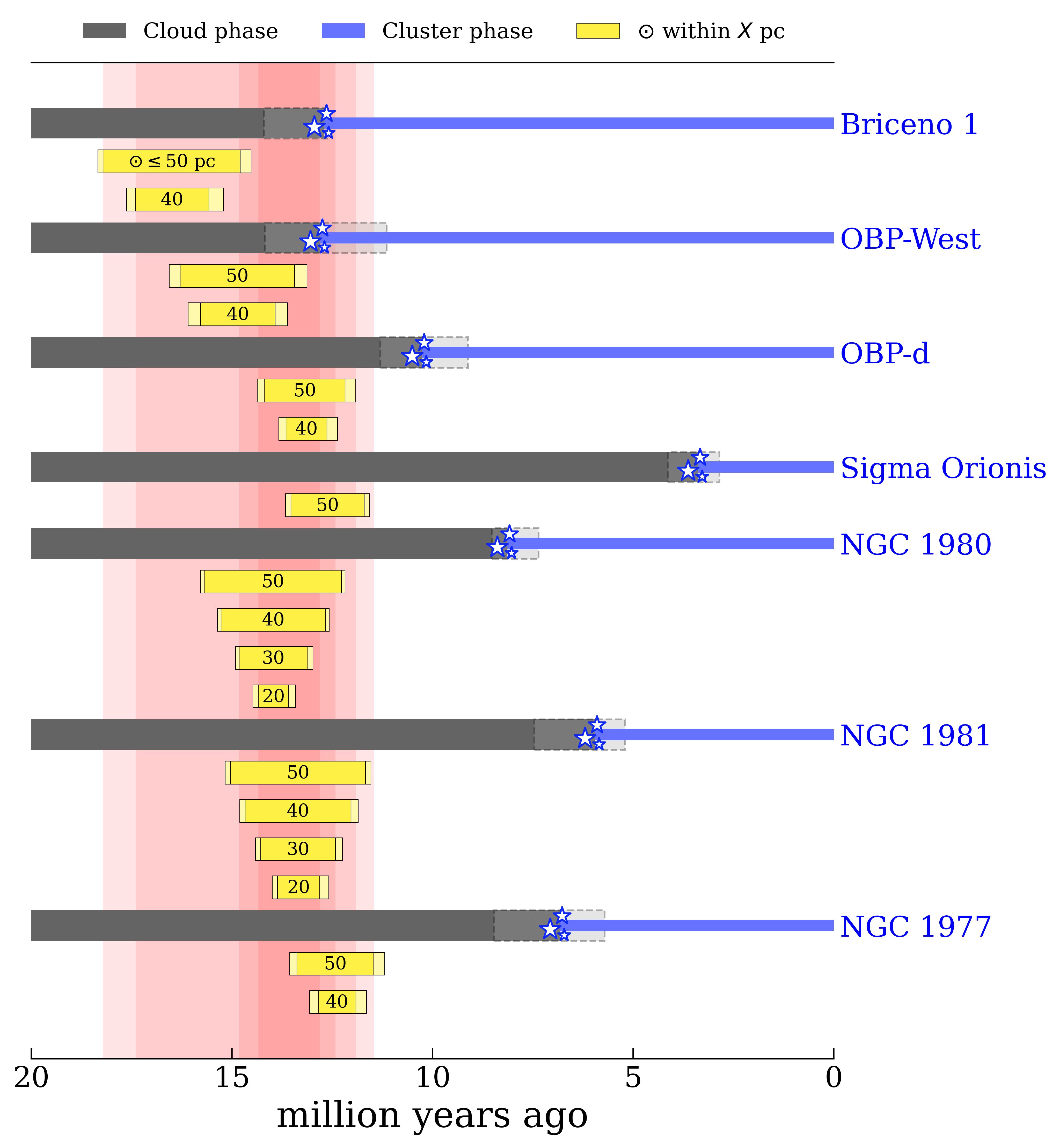}
    \caption{Significant encounters between the Solar System and the clusters of the Radcliffe wave during their cloud phase in the past 30 Myr, considering various threshold distances ($d_{\mathrm{Sun-cloud}}$; 50, 40, 30, and 20 pc). For each cluster, the cloud phase is represented by a horizontal gray band, the moment at which the cluster is formed is highlighted by three small stars, and the time period during which the cluster is fully formed (gas free phase) is denoted by a blue band. The age range of the cluster, as computed in this work, is indicated by a light-gray band with a dashed edge. The yellow bands highlight the time period during which the Solar System is within a certain distance from the clouds. Light-yellow represents the \correction{statistical} uncertainty of the crossing times, as computed from the tracebacks. The vertical red stripes summarize the time range during which the Solar System is passing through the gaseous part of the Radcliffe wave. The closer the transit, the redder the vertical stripe. The numerical equivalent of this plot is reported in Table~\ref{tab:Sun-RWclouds-interaction-times}.}
    \label{fig:Sun-RWclouds-interaction-times}
\end{figure}

\begin{table*}
	\centering
    \renewcommand{\arraystretch}{1.2}
    \caption{Time ranges and \correction{statistical} errors for significant encounters (crossing probability greater than 50\%) between the Sun and the Radcliffe wave’s clusters, assuming different threshold distances ($d_{\mathrm{Sun-cloud}}$).}
    \resizebox{0.85\textwidth}{!}{
    \setlength{\tabcolsep}{8pt}
	\begin{tabular}{l|cc|cc|cc|cc} 
        \hline
        \hline
        \vspace{0.08cm}
		 & \multicolumn{2}{|c|}{$d_{\mathrm{Sun-cloud}}\le50\,\mathrm{pc}$} &  \multicolumn{2}{|c|}{$d_{\mathrm{Sun-cloud}}\le40\,\mathrm{pc}$}  & \multicolumn{2}{|c|}{$d_{\mathrm{Sun-cloud}}\le30\,\mathrm{pc}$} & \multicolumn{2}{|c}{$d_{\mathrm{Sun-cloud}}\le20\,\mathrm{pc}$} \\
        \cline{2-3} \cline{4-5} \cline{6-7} \cline{8-9}    
        Name & $t_{\mathrm{enter}}$ & $t_{\mathrm{exit}}$ & $t_{\mathrm{enter}}$ & $t_{\mathrm{exit}}$ & $t_{\mathrm{enter}}$ & $t_{\mathrm{exit}}$ & $t_{\mathrm{enter}}$ & $t_{\mathrm{exit}}$ \\
         & $\mathrm{[Myr]}$ & $\mathrm{[Myr]}$ & $\mathrm{[Myr]}$ & $\mathrm{[Myr]}$ & $\mathrm{[Myr]}$ & $\mathrm{[Myr]}$ & $\mathrm{[Myr]}$ & $\mathrm{[Myr]}$ \\ 
		\hline 
        Briceno\,1 & $-18.2\pm0.1$ & $-14.8\pm0.3$ & $-17.4\pm0.2$ & $-15.6\pm0.4$ & - & - & - & - \\
        OBP-West & $-16.3\pm0.3$ & $-13.4\pm0.3$ & $-15.8\pm0.3$ & $-13.9\pm0.3$ & - & - & - & - \\
        OBP-d & $-14.2\pm0.2$ & $-12.2\pm0.3$ & $-13.6\pm0.2$ & $-12.6\pm0.3$ & - & - & - & - \\
        Sigma\,Orionis & $-13.5\pm0.1$ & $-11.7\pm0.1$ & - & - & - & - & - & - \\
        NGC\,1980 & $-15.7\pm0.1$ & $-12.3\pm0.1$ & $-15.3\pm0.1$ & $-12.7\pm0.1$ & $-14.8\pm0.1$ & $-13.1\pm0.1$ & $-14.3\pm0.1$ & $-13.6\pm0.2$ \\
        NGC\,1981 & $-15.0\pm0.1$ & $-11.7\pm0.1$ & $-14.7\pm0.1$ & $-12.0\pm0.2$ & $-14.3\pm0.1$ & $-12.4\pm0.2$ & $-13.9\pm0.1$ & $-12.8\pm0.2$ \\
        NGC\,1977 & $-13.4\pm0.2$ & $-11.5\pm0.3$ & $-12.8\pm0.2$ & $-11.9\pm0.3$ & - & - & - & - \\
        \hline
        Radcliffe wave & $-18.2\pm0.1$ & $-11.5\pm0.3$ & $-17.4\pm0.2$ & $-11.9\pm0.3$ & $-14.8\pm0.1$ & $-12.4\pm0.2$ & $-14.3\pm0.1$ & $-12.8\pm0.2$ \\
		\hline
	\end{tabular}}
    \tablefoot{This table serves as the numerical complement to Fig.~\ref{fig:Sun-RWclouds-interaction-times}.}
	\label{tab:Sun-RWclouds-interaction-times}
\end{table*}

\begin{figure*}
    \centering
    \includegraphics[width=1.0\textwidth]{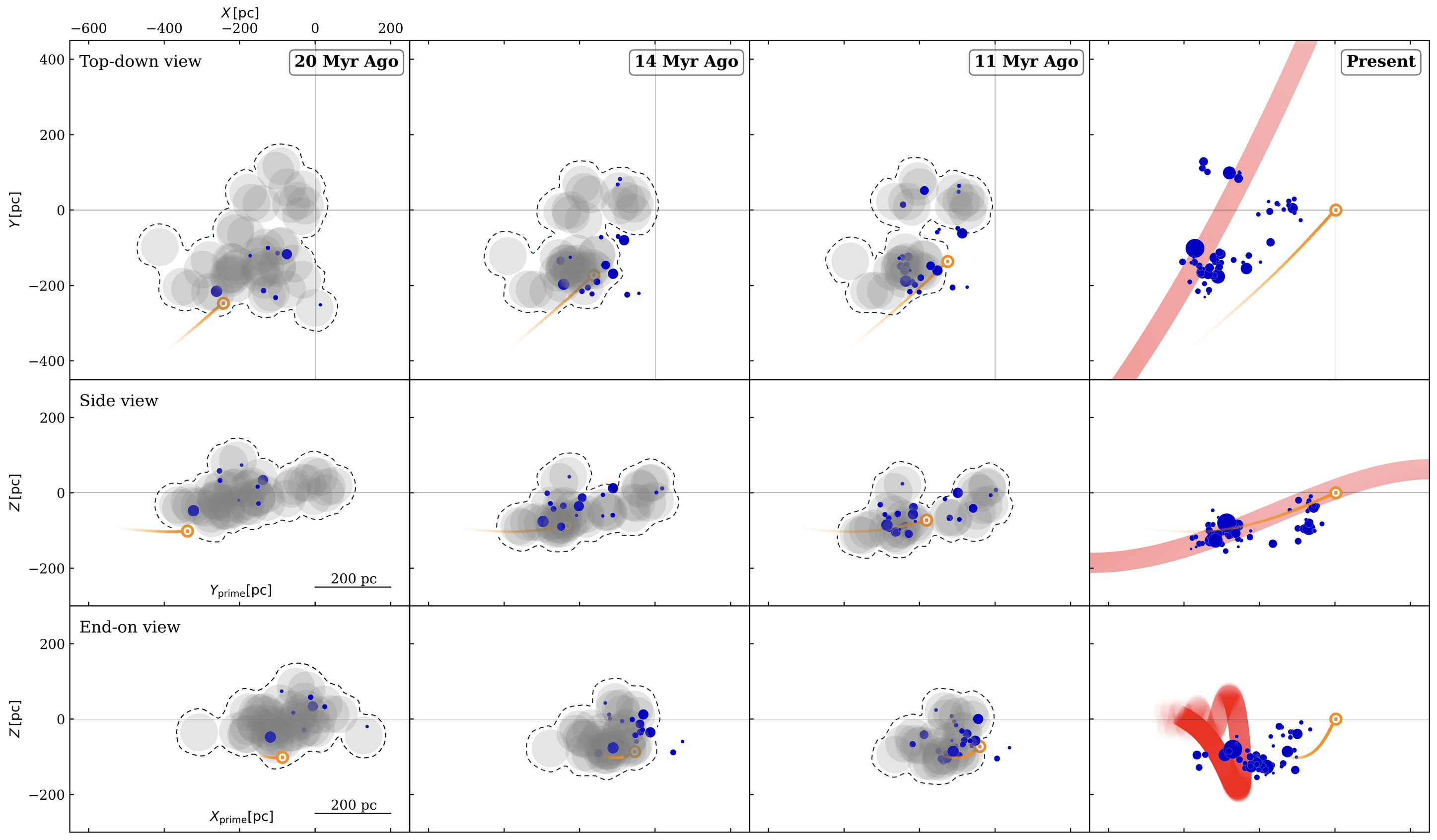}
    \caption{Selected time snapshots for the tracebacks of the orbits of the Sun, selected clusters, and relative parental clouds. In the columns from left to right, we depict the Sun approaching \correction{the} Radcliffe wave ($t$ = 20 Myr ago), the Sun within the gas of the Radcliffe wave ($t$ = 14 Myr ago), the Sun after it exited this gas structure ($t$ = 11 Myr ago), and the present day situation ($t$ = Present), respectively. The panels in the first row illustrate a top-down projection, while those in the second and third rows show side and end-on views, respectively, as depicted in Fig.~\ref{fig:RWoverview}. The Sun is denoted by a golden-yellow $\odot$, and its trail is represented by dotted points with decreasing opacity. The analyzed clusters are indicated as blue circles, whose sizes are proportional to the number of their stellar members. Clusters prior to their birth are represented by light-gray circles, symbolizing 50 pc radius gas clouds. \correction{We enclosed the clouds with a dashed black line to highlight the larger gas complex they are part of.} The best-fit model \citep{Konietzka2024} of the Radcliffe wave \correction{is shown as a light-red band in the present day panels}. An \href{https://www.aanda.org/articles/aa/olm/2025/02/aa52061-24/aa52061-24.html}{online} 3D animation is available \href{https://efremmaconi.github.io/emaconi_webpage/Sun_RW_crossing_3D_animation.html}{here}, viewable from any angle and at any time step over the past 30 Myr.}
    \label{fig:time_snapshots_tracebacks}
\end{figure*}

\section{Results and discussion}\label{sec:Results_Discussion}

\subsection{Solar System's crossing of the Radcliffe wave}\label{sec:ResDisc_crossing}

To assess a potential crossing of the Radcliffe wave by the Solar System, we computed, at each time step of the orbits tracebacks, the distances between the Sun and the clusters. For time intervals preceding the age of a given cluster, we considered its pre-birth trajectory as a first approximation of the orbit of the \correction{primordial} cloud associated with it. \correction{As described in Sect.~\ref{sec:RW_clusters_identification}, we interpret the motion of the analyzed clouds as tracing the motion of the larger gas complex they belong to.}
By computing the time ranges in which the Sun and the center of the parental gas clouds are closer than their radii —considered as threshold distances— we were able to determine when the Sun most likely \correction{crossed these clouds and} consequently the Radcliffe wave. We deemed a crossing significant if its probability of occurrence exceeds 50\%. This was computed by repeating the tracebacks of the orbits multiple times, as described in Sect.~\ref{sec:orbits_traceback}. 

Remarkably, we find that the past trajectories of the Solar System closely approached ($d_{\mathrm{Sun-cloud}}$ within 50 pc) certain selected clusters while they were in their cloud phase, hinting at a probable encounter between the Sun and the gaseous component of the Radcliffe wave. 
When considering a cloud as a sphere of gas spanning 50 pc, we observe that the Sun's orbit \correction{was} concurrently passing through multiple parental clouds associated with the clusters Briceno 1, OBP-West, OBP-d, Sigma Orionis, NGC 1980, NGC 1981, and NGC 1977 between  $18.2 \pm 0.1$ and $11.5 \pm 0.3$ Myr ago. These clusters are currently located within the \correction{Orion star-forming complex}. It is relevant and interesting to note, from a historical perspective, that a possible crossing of the Orion region by the Solar System was already suggested by \cite{Shapley1921}, based on much less reliable data.
Assuming a threshold distance of 40 pc, the Sun would still cross the gas clouds of all these clusters, with the exception of Sigma Orionis, approximately from $17.4 \pm 0.2 $ to $11.9 \pm 0.3$ Myr ago.
Considering 30 pc and 20 pc radii, the Solar System is within the parental clouds of NGC 1980 and NGC 1981 between $14.8 \pm 0.1$ and $12.4 \pm 0.2$ Myr ago, and $14.3 \pm 0.1$ and $12.8 \pm 0.2$ Myr ago, respectively.
These time ranges are approximate since gravitational scattering from the clouds and stellar feedback are not accounted for; however, \correction{given the short integration times \citep[less than 30 Myr; see e.g.,][]{Kamdar2021}, the presented approximations are a valid first step to better understand the past Sun and clouds interactions.}

The crossing time ranges of the Radcliffe wave by the Sun are shown in Fig.~\ref{fig:Sun-RWclouds-interaction-times} and listed in Table~\ref{tab:Sun-RWclouds-interaction-times}.
A complete animation illustrating the orbital trajectories of the clusters, clouds, and Sun over the past 30 Myr is presented in Fig.~\ref{fig:time_snapshots_tracebacks} (\href{https://efremmaconi.github.io/emaconi_webpage/Sun_RW_crossing_3D_animation.html}{interactive}). In the static version, we depict four different snapshots at $-20$ Myr, $-14$ Myr, $-11$ Myr, and the present. \correction{It can be seen that the Sun was} approaching, crossing, and leaving the Radcliffe wave. In Fig.~\ref{fig:time_snapshots_tracebacks}, we enclose the represented clouds with a dashed line, emphasizing the fact that they should be considered as part of a larger gas complex.
From the interactive version of the figure, it is possible to note that the orbits of these clusters exhibit common motion. For some, their birthplaces likely indicate a shared formation history, \correctionII{although they do not need to come from exactly the same point in space.
This is supported by the fact that some of the analyzed clusters  belong to one of the three families of clusters identified in the work by \citet{Swiggum2024}. For example, UPK 398, ASCC 18, ASCC 20, OCSN 64, OCSN 65, and CWNU 1072 are part of the Collinder 135 cluster family. Clusters in each of these three families converge toward each other when traced backward in time, consistent with shared formation origins \citep[see][Extended Data Fig.~1]{Swiggum2024}.
Additionally, previous} studies have already examined most of these clusters in the context of Orion \citep[see e.g.,][]{Bally1987,Brown1994,Briceno2008,Alves2012,Chen2020b,Kounkel2020,Grossschedl2021}.


For the seven clusters to which the Sun's orbits get closer than 50 pc with a high probability, the estimated radii of the densest parts of the associated gas clouds (densities higher than \correction{40 particles} per $\mathrm{cm}^3$) range between \correction{11.9–21.2 pc} when considering 1\% SFE, or \correction{8.2–14.5 pc} for a 3\% SFE (refer to Sect.~\ref{sec:gas_cloud_properties} and Table~\ref{tab:RWclusters-properties}). 
Especially noteworthy is the case of NGC 1980, one of the two clusters to which the Solar System approaches within 20 pc. For this cluster, we estimated a radius of \correction{21.2 pc (14.5 pc)}, which further supports our crossing hypothesis. It is important to clarify that the estimated radii pertain to the densest part of the clouds, which can then be surrounded by tenuous gas enveloping the central regions \citep[see e.g.,][]{Snow2006}. Moreover, these radii should be considered as lower limits, given the likely incompleteness of the cluster catalog, as previously highlighted in Sect.~\ref{sec:gas_cloud_properties}.
We remark that these findings hold true even when assuming other initial conditions for the Sun’s parameters and different Milky Way potentials (see Appendix~\ref{app:initialCondTest} for details). 

\begin{figure*}
    \centering
    \includegraphics[width=1.0\textwidth]{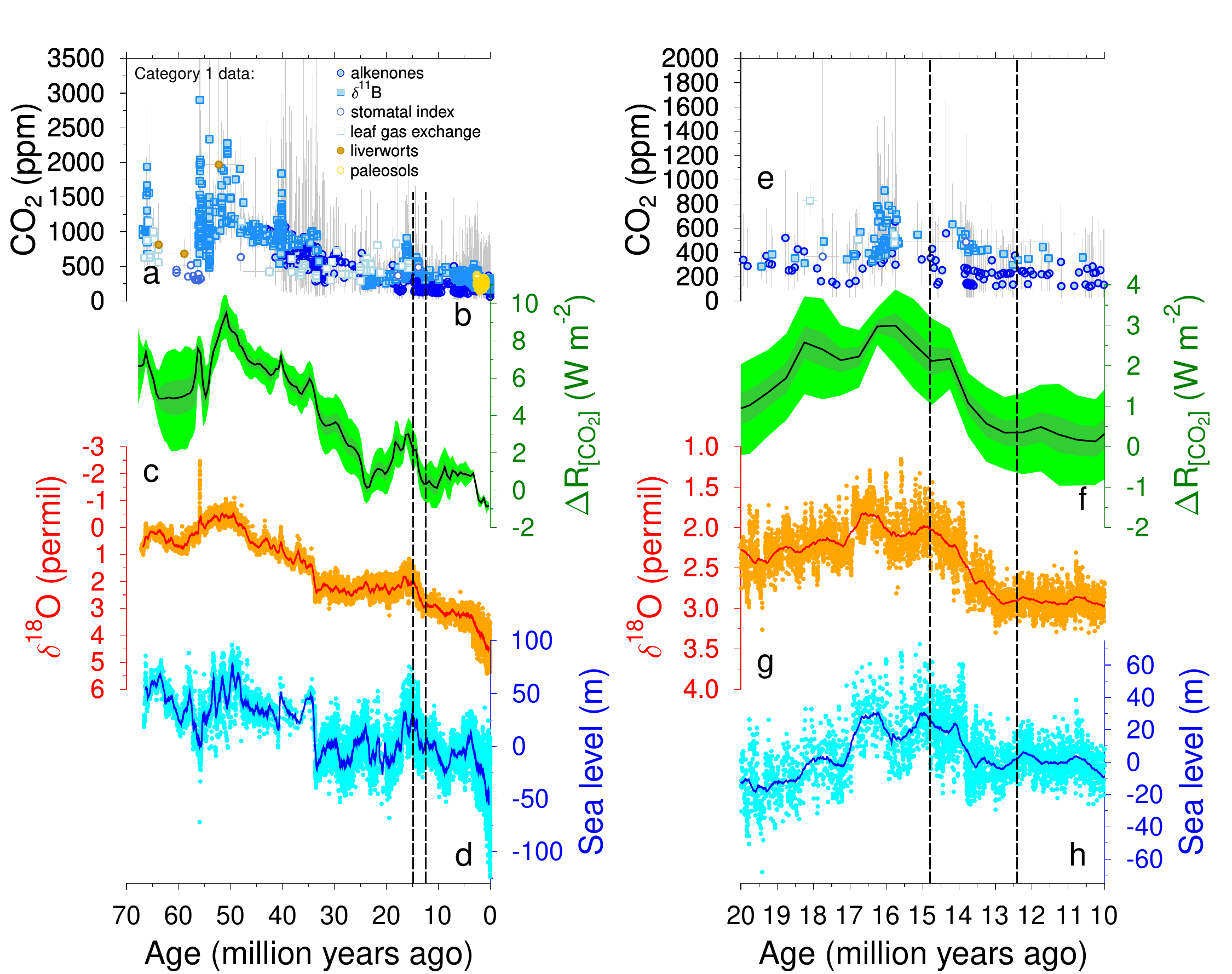}
    \caption{Cenozoic climate change. (a) Most recent compilation of category 1 (most trustworthy) atmospheric $\mathrm{CO}_2$ data \citep{CenCO2PIP2023} as a function of proxy with their fully developed uncertainty estimates (95\% confidence intervals). (b) $\mathrm{CO}_2$ radiative forcing ($\Delta\mathrm{R}_{[\mathrm{CO}_2]} = 5.35 [\ln(\mathrm{CO}_2) - \ln(\mathrm{CO}_{2,0})] \, \mathrm{W}\,\mathrm{m}^{-2}$, $\mathrm{CO}_{2,0}$ = 278 ppm) of a 500-kyr running mean through data with median and 50 and 95\% credible intervals: dark and light-green shading, respectively \citep{CenCO2PIP2023}. (c) benthic $\delta^{18}\mathrm{O}$ recording a mixture of deep ocean temperature and sea level, raw data (points) and 500-kyr running mean (line) \citep{Westerhold2020}. (d) Reconstructed sea level change, raw data (points) and 500-kyr running mean (line) \citep{Miller2020}. Vertical broken lines mark the most likely time window ($14.8-12.4$ Myr ago, for $d_{\mathrm{Sun-cloud}}$ within $20-30$ pc) of the Solar System being located in the dense region of the ISM suggested by this study. (e-h) Zoom in on the time window $20-10$ Myr ago.}
    \label{fig:cenozoic_climate_change}
\end{figure*}

\subsection{Interdisciplinary bridges: Potential geological and climate evidences}

The crossing of a dense region of the ISM by the Sun, such as a gas cloud or a SN blast wave, can impact the Solar System in various ways \citep[see e.g.,][]{Fields2023,Opher2024}. For example, \correction{the compression of the heliosphere by enhanced ram pressure exposes parts of the Solar System to the cold and dense ISM} \citep[see e.g.,][]{Miller2022,Opher2024,Opher2024b}. The amount of dust loaded into Earth's atmosphere would also increase, probably enhancing the delivery of radioisotopes (e.g., $^{60}$Fe) via dust grains \citep[see e.g.,][]{Altobelli2005,Breitschwerdt2016}. \correction{This could lead to anomalies in geological radionuclide records} \citep{koll2019,Wallner2015,Wallner2021} and could provide evidence of the passage of the Sun through the Radcliffe wave. 
Our estimates suggest that the Orion region traversed by the Sun may have been enriched with radioisotopes from $3^{+1}_{-2}$ SN before 11.5 Myr ago, with the potential incompleteness \correction{of the stellar membership in the catalogs} increasing this estimate. For the estimation of the number of past SN events, we refer to Appendix~\ref{app:SN_computation}. Although current $^{60}$Fe data do not cover our period of interest \citep{Fields2023}, future instrumentation is expected to be sensitive enough to analyze this time period.

Furthermore, an increased amount of dust could impact Earth’s radiation budget, potentially leading to a cooling effect \citep[see e.g.,][]{Talbot1977,Pavlov2005}.
Notably, our estimated time interval for the Solar System's potential location within a dense ISM region \correction{(about \mbox{$14.8-12.4$ Myr ago} for a distance of $20-30$ pc from the center of a gas cloud)} overlaps with the Middle Miocene climate transition \citep{Steinthorsdottir2021}. During this period, the expansion of the Antarctic ice sheet \citep{Miller2020} and global cooling \citep{Westerhold2020} marked Earth's final transition to persistent large-scale continental glaciation in Antarctica (see Fig.~\ref{fig:cenozoic_climate_change}, panel c-d and g-h). The ice sheet-climate interactions during the Miocene are complex \citep{Knorr2014,Stap2024} and the evolving understanding suggests that this cooling phase was possibly caused by falling atmospheric $\mathrm{CO}_2$ concentrations \citep{CenCO2PIP2023} (Fig.~\ref{fig:cenozoic_climate_change}, panel b and f). However, $\mathrm{CO}_2$ reconstructions beyond what is covered in Antarctic ice core data (the last 0.8 Myr) are highly uncertain (Fig.~\ref{fig:cenozoic_climate_change}, panel a and e) and $\mathrm{CO}_2$, when reconstructed from alkenones \citep{vandeWal2011} (Fig.~\ref{fig:cenozoic_climate_change}, panel e, light blue circles), seem indeed to suggest that the Middle Miocene cooling was not directly coupled to $\mathrm{CO}_2$ radiative forcing. Therefore, it is intriguing to consider that the passage of our Solar System through a dense region of the ISM might have contributed to this climate transition, even if this remains speculative and currently lacks direct proof. We cautiously point out such possibilities to make the community aware of maybe overlooked processes, but we are conscious that available $^{60}$Fe data do not extend beyond 10 Myr ago \citep{Fields2023} and to our knowledge no rise in dust load has yet been discovered around 14 Myr ago in deep sea sediments \citep{Rea1994}. 
\correction{Furthermore, the suggested decoupling of Middle Miocene temperature and alkenone-based $\mathrm{CO}_2$, that opens up the possibility that not $\mathrm{CO}_2$ but other processes are responsible for the reconstructed cooling, needs to be taken with caution, since recently various authors have suggested that some fundamental difficulties exist in studies that used alkenones for the reconstruction of atmospheric $\mathrm{CO}_2$ \citep[e.g.,][]{Phelps2021,Rae2021,Brandenburg2022}.}

\correction{We compared the radiative forcing and climate response during the Middle Miocene cooling period to what is known for the ice ages of the Pleistocene ($2.6-0.01$ Myr ago), in order to estimate by how much the extraterrestrial dust flux to Earth needs to have changed during the Solar System’s crossing of the Radcliffe wave to serve as the primary driver of this climate transition - alternatively to atmospheric $\mathrm{CO}_2$ concentrations.
We based our analysis on the Pleistocene, as this period has a significantly more extensive data coverage. However, we acknowledge that this is an imperfect comparison, partly due to differences in the time scales of climate change (Myr in the Middle Miocene versus $10-100$ kyr in the Pleistocene), and because the $\mathrm{CO}_2$ radiative forcing changes during the Middle Miocene climate transition and late Pleistocene ice ages are of similar size, yet linked to different climate responses.
Nonetheless, we used this comparison to obtain an order of magnitude estimate of the required change in dust flux.

During the Middle Miocene climate transition, the long-term mean radiative forcing of $\mathrm{CO}_2$ shows a reduction by \mbox{$\sim2\, \mathrm{W} \, \mathrm{m}^{-2}$} (Fig.~\ref{fig:cenozoic_climate_change}, panel f). Concurrently, the long-term and global mean surface temperature, estimated from benthic $\delta^{18}\mathrm{O}$ (Fig.~\ref{fig:cenozoic_climate_change}, panel g), decreased by more than $2 \, \mathrm{K}$ \citep{Westerhold2020} and sea level dropped by $\sim20 \, \mathrm{m}$ (Fig.~\ref{fig:cenozoic_climate_change}, panel h). We refer to \cite{Rohling2024} for the ongoing discussion on apparent discrepancies in reconstructed Cenozoic climate change based on different proxies. 
During glacial periods of the late Pleistocene, a similar reduction in $\mathrm{CO}_2$ radiative forcing, together with other forcing and feedback processes \citep{Koehler2010}, led to a global mean surface cooling of approximately $6 \, \mathrm{K}$ \citep{Tierney2020,Clark2024} and to land ice sheets growth, mainly in North America and Eurasia. This ice sheets expansion corresponded to a sea level drop of about $\sim120 \, \mathrm{m}$ \citep{Gowan2021}.
The root cause of Pleistocene glaciations is understood to be the changes of Earth's orbital parameters \citep{Milankovic1941,Barker2024}, which led to variations in incoming solar radiation \citep{Laskar2004}. Nevertheless, reduced greenhouse gas concentrations, increased atmospheric dust load, and higher surface albedo are important contributions necessary to drive Earth's climate into an ice age \citep{Koehler2010}.

During the late Pleistocene ice ages, the glacial dust (i.e., global dust during ice ages) deposition rate was $2-4$ times larger than today \citep{Albani2016, Mahowald2023} and, although the impact of dust on the climate is complex \citep{Kok2023} and with significant uncertainties \citep{Mahowald2023}, it contributed to a radiative forcing of about $-1\, \mathrm{W} \, \mathrm{m}^{-2}$ \citep{Koehler2010,Shaffer2018,Sherwood2020}, roughly half of the change in radiative forcing caused by atmospheric $\mathrm{CO}_2$.
Given that the present-day global dust deposition rate on Earth is $5 \cdot 10^{15} \, \mathrm{g} \, \mathrm{yr}^{-1}$ \citep{Kok2021}, it would need to rise to $(2-4) \cdot 10^{16} \, \mathrm{g} \, \mathrm{yr}^{-1}$ in order achieve a radiative forcing similar to that proposed so far for $\mathrm{CO}_2$ during the Middle Miocene climate transition. 
Furthermore, the current incoming flux of extraterrestrial dust load is $(1-2) \cdot 10^{10} \, \mathrm{g} \, \mathrm{yr}^{-1}$ on top of the atmosphere, which decreases to $(4-7) \cdot 10^{9} \, \mathrm{g} \, \mathrm{yr}^{-1}$ at the surface \citep[][]{Love1993,Plane2012,Rojas2021}. Therefore, for the crossing of the Radcliffe wave to be the main driver of the Middle Miocene cooling, the extraterrestrial dust flux would need to rise by about $6-7$ orders of magnitude to produce such a large anomaly in the radiation budget and the resulting climate effects.
At present, the Sun is located at the edge of the Local Interstellar Cloud (LIC), a low-density cloudlet with $n_{\mathrm{H}}\sim 0.03-0.1\,\mathrm{cm}^{-3}$ \citep[see e.g.,][]{Gry2014}, located within a SN-generated hot void known as the Local Bubble, which has a characteristic density of $\sim 0.01\,\mathrm{cm}^{-3}$ \citep[see e.g.,][]{coxANDreynolds1987,Linsky2021,Zucker2022,ONeill2024}. As the ISM density of the Radcliffe wave is significantly higher (ranging from $10^1$ to  $10^3\,\mathrm{cm}^{-3}$) than the region currently traversed by the Solar System, the increase in the dust load on Earth could be at least of $2-5$ orders of magnitude. 
Consequently, considering all factors and disregarding other differences in the climate system between the late Pleistocene and the Middle Miocene, our study suggests that the potential rise in extraterrestrial dust during the Solar System's crossing of the Radcliffe wave may have been $1-5$ orders of magnitude smaller than necessary to fully account for the Middle Miocene climate transition as observed in the geological record. 
At the moment, we can therefore infer that this process likely played a limited role in the Middle Miocene climate transition. However, once some of the underlying assumptions of this estimate are better constrained, the contribution to this event might be reassessed in either direction.
Additionally, due to the nonpermanent nature of any extraterrestrial dust influx associated with the Solar System’s crossing, this process alone is unlikely to account for the long-term effect of a reduction in atmospheric $\mathrm{CO}_2$. While it may have influenced climate during the multi-million years duration of the passage, other forcings or feedbacks would be required to explain the persistence of low temperatures and sea levels following the Middle Miocene climate transition.
}

To conclude, present day knowledge suggests that beyond small solar-driven climate oscillations \citep{Eddy1976}, the long-term energy output of the Sun \citep{Gough1981} together with Earth’s orbital parameters \citep{Milankovic1941}, plate tectonics \citep{Scotese2021}, deep carbon cycle \citep{Muller2022}, internal feedback \citep{Ganopolski2024}, and very few large-scale meteorite impacts \citep{Osinski2022,Zorzi2022} can explain a wide spectrum of the reconstructed changes and variability in Cenozoic climate. So far, all proposed additional extra-terrestrial influences have, to our knowledge, remained in a hypothetical state \citep{Pavlov2005,Opher2024} or even been discarded \citep{Berger1999,Carslaw2002,Bard2006}. However, we cannot rule out the possibility that an extraordinary amount of dust in the entire inner Solar System might have led to a reduction of incoming Solar radiation and a cooling on Earth, similar to what has been proposed for the triggering of the mid-Ordovician ice age 466 My ago \citep{Schmitz2019}. \correction{Furthermore, as shown by \cite{Miller2024}, some effects of the Solar System’s passage through a dense ISM region may be linked to nonpermanent, seasonal variations in cloud formation on Earth, which are more difficult to detect in paleo-records.}

\correction{
\subsection{Caveats}
Our results are based on the tracebacks of the orbits of the Solar System and of the clusters associated with the Radcliffe wave. As noted throughout the text, this method requires some approximations due to inherent difficulties in modeling the past structure and evolution of the gas. For example, we simplified the diverse and complex morphologies of the molecular clouds by assuming a spherical shape. Based on the principle of momentum conservation, we assumed that the motion of young clusters still reflects the movement of the gas nurseries from which they formed. We acknowledge that gravitational interactions and feedback from massive stars have likely influenced parts of the gas clouds and, consequently, the velocities of the clusters which formed within them. Thus, our tracebacks should be considered as a first approximation of the actual orbits. With these assumptions, we showed the general patterns of motion and estimate a time window during which the Solar System may have crossed the Radcliffe wave. Future studies will need to examine in greater detail the possible effects of gravity and feedback-induced displacements.
}

\section{Conclusions}\label{sec:Conclusions}

Our investigation reveals that the Solar System likely passed through the Orion region of the Radcliffe wave gas structure. By tracing back the orbits of the Sun and the Radcliffe wave’s clusters, we constrain the time range of this crossing to be between $18.2 \pm 0.1$ and $11.5 \pm 0.3$ Myr ago, with the closest approaches ($d_{\mathrm{Sun-cloud}}$ within $20-30$ pc) occurring within the interval of $14.8 \pm 0.1$ to $12.4 \pm 0.2$ Myr ago. 
\correction{As we do not account for potential gravitational interactions and feedback forces, we consider these time ranges to be preliminary approximations. Our results remain consistent even when varying solar parameters and Milky Way gravitational potentials.

The potentially increased amount of dust in the inner Solar System and in Earth’s atmosphere resulting from such an interaction provides a framework for searching for isotopic anomalies in geological records older than 10 Myr. 
This is supported by our estimate that approximately $3_{-2}^{+1}$ SN occurred in the region traversed by the Sun, thus seeding it with freshly produced isotopes, such as $^{60}$Fe.
Additionally, we find that the period of closest approach was synchronous with the reorganization of the Earth’s climate known as the Middle Miocene climate transition, even though a causal connection between the two events remains speculative and lacks direct evidence.
In conclusion, our investigations underline} the importance of studying the Galactic environment encountered by the Solar System during its orbit, along with the potential effects this may have on Earth.

\section*{Data availability}
The supplementary figures of the Appendix are available online via Zenodo at the following link: \href{https://doi.org/10.5281/zenodo.14626660}{https://doi.org/10.5281/zenodo.14626660}. The code used for the analysis will be shared by EM upon reasonable request. 

\begin{acknowledgements}
\correction{We thank the anonymous referee for the insightful comments, which have improved the quality and readability of the paper.}
JA and EM were co-funded by the European Union (ERC, ISM-FLOW, 101055318). \correction{S.R. acknowledges funding by the Austrian Research Promotion Agency (FFG, \url{https://www.ffg.at/}) under project number FO999892674.} \correction{JG gratefully acknowledges co-funding from the European Union, the Central Bohemian Region, and the Czech Academy of Sciences, as part of the MERIT fellowship (MSCA-COFUND Horizon Europe, Grant Agreement No.~101081195). JG acknowledges the Collaborative Research Center 1601 (SFB 1601) funded by the Deutsche Forschungsgemeinschaft (DFG, German Research Foundation) – 500700252.}
This work has made use of data from the European Space Agency (ESA) mission \textit{Gaia} \href{https://www.cosmos.esa.int/gaia}{https://www.cosmos.esa.int/gaia}, processed by the \textit{Gaia} Data Processing and Analysis Consortium (DPAC,
\href{https://www.cosmos.esa.int/web/gaia/dpac/consortium}{https://www.cosmos.esa.int/web/gaia/dpac/consortium}). Funding for the DPAC has been provided by national institutions, in particular the institutions participating in the \textit{Gaia} Multilateral Agreement.
This research has made use of publicly available software libraries, including galpy \citep{Bovy2015}, Astropy \citep{astropy}, NumPy \citep{numpy}, ALADIN \citep{aladin}, TOPCAT \citep{topcat}, matplotlib \citep{matplotlib}, and plotly.
\correction{The IMF has been sampled using the python code available at the following repository on github: \href{https://github.com/keflavich/imf}{https://github.com/keflavich/imf}.
The ages of the clusters have been estimated using the {\tt Chronos} python code available at the following repository on github: \href{https://github.com/ratzenboe/Chronos}{https://github.com/ratzenboe/Chronos}.}

\end{acknowledgements}

%
%
\bibliographystyle{aa} 
\bibliography{references} 

\begin{appendix} 

\section{Properties of the clusters}

\begin{table*}[h!]
	\centering
    \renewcommand{\arraystretch}{1.1}
	\caption{Cartesian coordinates ($X,\,Y,\,Z$) and velocities ($U,\,V,\,W$) with respect to the Sun, together with the corresponding errors, for the 56 identified clusters of the Radcliffe wave within the region of interest for this study.}
    \resizebox{0.90\textwidth}{!}{
	\begin{tabular}{lcccccccccccc} 
        \hline
        \hline
        \vspace{0.08cm}
		Name in catalog & $X$ & $Y$ & $Z$ & $U$ &$V$ & $W$ & $X_{\mathrm{err}}$ & $Y_{\mathrm{err}}$ & $Z_{\mathrm{err}}$ & $U_{\mathrm{err}}$ &$V_{\mathrm{err}}$ & $W_{\mathrm{err}}$ \\
  		 & $\mathrm{[pc]}$ & $\mathrm{[pc]}$ & $\mathrm{[pc]}$ & $\mathrm{[km\,s^{-1}]}$ &$\mathrm{[km\,s^{-1}]}$ & $\mathrm{[km\,s^{-1}]}$ & $\mathrm{[pc]}$ & $\mathrm{[pc]}$ & $\mathrm{[pc]}$ & $\mathrm{[km\,s^{-1}]}$ &$\mathrm{[km\,s^{-1}]}$ & $\mathrm{[km\,s^{-1}]}$ \\
		\hline 
        Briceno\,1 & -302.44 & -116.94 & -107.91 & -18.19 & -8.26 & -4.60 & 0.49 & 0.20 & 0.20 & 0.21 & 0.08 & 0.08 \\
        ASCC\,18 & -359.37 & -147.74 & -130.29 & -25.13 & -8.07 & -7.05 & 2.21 & 1.03 & 0.93 & 0.49 & 0.27 & 0.21 \\
        Theia\,13 & -351.87 & -165.40 & -127.41 & -25.92 & -12.70 & -9.39 & 0.77 & 0.39 & 0.31 & 0.39 & 0.17 & 0.16 \\
        Sigma\,Orionis & -336.25 & -169.96 & -116.64 & -25.54 & -15.57 & -6.86 & 0.57 & 0.32 & 0.25 & 0.18 & 0.14 & 0.12 \\
        NGC\,1980 & -309.96 & -175.65 & -127.38 & -22.22 & -12.84 & -6.59 & 0.50 & 0.27 & 0.20 & 0.16 & 0.09 & 0.07 \\
        UBC\,207 & -319.00 & -171.44 & -124.37 & -24.13 & -12.60 & -7.13 & 1.11 & 0.57 & 0.40 & 0.33 & 0.20 & 0.14 \\
        NGC\,1977 & -321.63 & -174.10 & -127.01 & -24.49 & -15.65 & -8.40 & 0.70 & 0.38 & 0.29 & 0.43 & 0.24 & 0.20 \\
        \hline
        ASCC\,19 & -301.08 & -139.19 & -116.56 & -18.49 & -11.54 & -6.62 & 0.50 & 0.34 & 0.31 & 0.79 & 0.37 & 0.32 \\
        ASCC\,20 & -318.64 & -126.23 & -109.39 & -27.19 & -8.93 & -9.29 & 1.02 & 0.39 & 0.33 & 0.31 & 0.13 & 0.11 \\
        ASCC\,21 & -305.74 & -111.92 & -96.40 & -17.52 & -8.76 & -3.63 & 0.37 & 0.37 & 0.18 & 0.66 & 0.24 & 0.21 \\
        Alessi-Teutsch\,10 & -351.13 & 111.51 & -129.20 & -20.21 & -4.87 & -9.93 & 1.91 & 0.64 & 0.88 & 1.58 & 0.57 & 0.52 \\
        CWNU\,1028 & -265.22 & -132.82 & -143.96 & -15.76 & -10.10 & -7.87 & 1.61 & 1.22 & 0.83 & 0.40 & 0.33 & 0.35 \\
        CWNU\,1072 & -362.98 & -215.24 & -121.07 & -24.92 & -11.66 & -12.68 & 2.94 & 1.93 & 0.98 & 0.99 & 0.63 & 0.36 \\
        CWNU\,1088 & -384.68 & -190.09 & -117.90 & -13.10 & -16.78 & -6.28 & 16.94 & 7.52 & 4.82 & 2.18 & 1.48 & 0.73 \\
        CWNU\,1092 & -380.77 & -139.44 & -47.32 & -23.97 & -13.06 & -12.87 & 2.13 & 0.99 & 0.69 & 0.75 & 0.10 & 0.14 \\
        CWNU\,1106 & -374.42 & -125.68 & -66.33 & -24.90 & -10.93 & -13.93 & 2.26 & 0.82 & 0.58 & 0.01 & 0.17 & 0.24 \\
        CWNU\,1129 & -150.26 & 15.18 & -43.44 & -17.07 & -13.10 & -6.93 & 0.61 & 1.23 & 0.34 & 0.25 & 0.11 & 0.10 \\
        Collinder\,69 & -370.78 & -101.25 & -81.02 & -24.46 & -11.19 & -5.60 & 0.48 & 0.29 & 0.19 & 0.15 & 0.10 & 0.06 \\
        HSC\,1250 & -252.15 & 97.91 & -100.59 & -16.80 & -9.42 & -7.20 & 1.01 & 0.91 & 0.99 & 0.66 & 0.40 & 0.39 \\
        HSC\,1262 & -348.18 & 128.73 & -96.31 & -23.82 & -2.09 & -7.16 & 0.92 & 0.45 & 0.36 & 0.68 & 0.31 & 0.22 \\
        HSC\,1318 & -122.48 & 23.79 & -34.52 & -16.45 & -11.98 & -10.88 & 0.34 & 0.31 & 0.23 & 0.28 & 0.13 & 0.10 \\
        HSC\,1340 & -112.14 & 5.26 & -39.44 & -13.82 & -6.32 & -9.99 & 0.56 & 0.83 & 0.50 & 0.52 & 0.08 & 0.19 \\
        HSC\,1373 & -107.86 & 29.33 & -82.42 & -12.43 & -5.41 & -5.95 & 1.43 & 2.48 & 0.81 & 0.91 & 0.29 & 0.99 \\
        HSC\,1481 & -203.09 & -10.93 & -46.30 & -16.40 & -8.05 & -8.83 & 1.56 & 2.51 & 0.77 & 1.54 & 0.21 & 0.09 \\
        HSC\,1633 & -310.87 & -155.53 & -136.51 & -17.98 & -12.91 & -5.71 & 0.93 & 0.55 & 0.48 & 1.10 & 0.44 & 0.36 \\
        HSC\,1640 & -170.63 & -85.40 & -135.32 & -6.83 & -8.63 & -6.01 & 2.30 & 1.40 & 1.16 & 0.88 & 0.36 & 0.56 \\
        HSC\,1648 & -228.05 & -120.40 & -117.82 & -10.75 & -11.54 & -7.14 & 1.40 & 0.51 & 0.53 & 1.19 & 0.67 & 0.65 \\
        HSC\,1653 & -242.87 & -138.68 & -126.54 & -12.48 & -7.97 & -6.28 & 0.96 & 0.57 & 0.54 & 1.50 & 0.86 & 0.72 \\
        HSC\,1692 & -197.27 & -137.93 & -101.25 & -22.99 & -5.87 & -6.85 & 3.75 & 2.58 & 1.65 & 0.65 & 0.30 & 0.48 \\
        IC\,348 & -279.68 & 99.15 & -95.34 & -16.38 & -5.93 & -7.42 & 4.47 & 1.56 & 1.56 & 0.14 & 0.08 & 0.07 \\
        L\,1641S & -333.57 & -212.08 & -135.95 & -16.74 & -11.52 & -7.12 & 1.92 & 1.20 & 0.81 & 0.44 & 0.27 & 0.19 \\
        Mamajek\,3 & -91.23 & -26.63 & -27.52 & -11.04 & -19.01 & -8.52 & 0.89 & 0.55 & 0.40 & 0.82 & 0.29 & 0.25 \\
        NGC\,1333 & -253.54 & 100.43 & -101.39 & -16.39 & -10.66 & -9.52 & 0.86 & 0.38 & 0.38 & 0.62 & 0.31 & 0.27 \\
        NGC\,2068 & -355.29 & -166.57 & -100.86 & -23.81 & -11.87 & -8.68 & 0.89 & 0.50 & 0.30 & 0.28 & 0.17 & 0.17 \\
        OC\,0340 & -352.44 & -160.61 & -113.16 & -24.00 & -11.95 & -9.57 & 3.31 & 1.54 & 1.09 & 1.96 & 0.74 & 0.50 \\
        OCSN\,50 & -175.67 & 22.91 & -71.36 & -14.35 & -5.76 & -5.32 & 2.04 & 1.63 & 0.53 & 0.96 & 0.16 & 0.43 \\
        OCSN\,56 & -371.51 & -138.14 & -108.89 & -28.20 & -8.28 & -9.37 & 1.16 & 0.80 & 0.67 & 1.00 & 0.36 & 0.39 \\
        OCSN\,59 & -318.14 & -138.39 & -155.33 & -18.37 & -10.62 & -7.70 & 0.62 & 0.40 & 0.57 & 1.19 & 0.63 & 0.50 \\
        OCSN\,61 & -331.92 & -152.78 & -110.51 & -26.86 & -12.36 & -11.18 & 0.81 & 0.45 & 0.41 & 0.83 & 0.39 & 0.29 \\
        OCSN\,64 & -268.51 & -132.13 & -122.54 & -26.17 & -5.31 & -5.58 & 14.02 & 6.97 & 6.04 & 1.31 & 0.78 & 0.68 \\
        OCSN\,65 & -356.47 & -170.22 & -125.64 & -25.48 & -9.35 & -9.48 & 1.23 & 0.67 & 0.52 & 1.08 & 0.48 & 0.46 \\
        OCSN\,68 & -345.31 & -194.74 & -135.39 & -25.48 & -9.74 & -11.14 & 1.52 & 1.55 & 0.80 & 0.84 & 0.48 & 0.35 \\
        OCSN\,70 & -334.34 & -221.18 & -144.60 & -14.34 & -12.41 & -6.35 & 0.78 & 0.60 & 0.48 & 0.44 & 0.28 & 0.30 \\
        OC\,0279 & -255.68 & 84.10 & -80.35 & -18.21 & -9.86 & -8.76 & 1.02 & 0.50 & 0.54 & 0.53 & 0.20 & 0.16 \\
        OC\,0280 & -337.87 & 101.55 & -95.04 & -20.82 & -5.01 & -7.35 & 1.30 & 0.46 & 0.77 & 1.17 & 0.34 & 0.35 \\
        OC\,0339 & -310.52 & -139.57 & -108.15 & -18.43 & -11.20 & -5.15 & 0.52 & 0.43 & 0.45 & 0.26 & 0.12 & 0.08 \\
        OC\,0356 & -344.73 & -230.65 & -149.48 & -13.86 & -13.13 & -6.99 & 0.85 & 0.67 & 0.45 & 0.82 & 0.27 & 0.12 \\
        Theia\,54 & -153.64 & 17.76 & -21.40 & -15.87 & -14.82 & -10.38 & 0.72 & 0.51 & 0.76 & 0.76 & 0.15 & 0.12 \\
        Theia\,65 & -107.30 & -6.84 & -9.24 & -12.61 & -18.73 & -8.63 & 0.52 & 0.74 & 0.66 & 0.75 & 0.13 & 0.11 \\
        Theia\,66 & -135.79 & 1.51 & -48.71 & -16.81 & -15.20 & -7.59 & 0.45 & 0.47 & 0.33 & 0.36 & 0.12 & 0.14 \\
        Theia\,7 & -122.75 & 12.78 & -34.07 & -16.02 & -11.06 & -9.48 & 0.69 & 0.43 & 0.25 & 0.26 & 0.11 & 0.14 \\
        Theia\,93 & -172.95 & -3.23 & -19.75 & -17.47 & -13.49 & -9.22 & 1.01 & 0.70 & 0.65 & 1.08 & 0.31 & 0.21 \\
        UBC\,17a & -306.62 & -150.75 & -103.93 & -17.99 & -12.63 & -4.93 & 0.57 & 0.53 & 0.39 & 0.37 & 0.24 & 0.14 \\
        UPK\,398 & -403.81 & -137.17 & -86.49 & -27.14 & -9.67 & -13.53 & 1.43 & 0.60 & 0.33 & 1.90 & 0.65 & 0.48 \\
        UPK\,402 & -356.17 & -153.64 & -84.64 & -20.06 & -11.08 & -8.23 & 1.49 & 0.67 & 0.41 & 1.06 & 0.41 & 0.30 \\
        UPK\,422 & -234.56 & -154.66 & -86.42 & -12.13 & -8.87 & -4.82 & 0.96 & 0.51 & 0.39 & 0.49 & 0.33 & 0.21 \\
        \hline
	\end{tabular}
    }
    \tablefoot{The first seven clusters are those with which the Sun's orbits get closer than 50 pc during their cloud phase, as identified in this study (\correction{see} Table~\ref{tab:Sun-RWclouds-interaction-times}).}
	\label{tab:RWclusters-6Dinfo-allClusters}
\end{table*}

\begin{table*}
	\centering
    \renewcommand{\arraystretch}{1.5}
    \caption{Properties of the 56 Radcliffe wave's clusters selected for this study.} 
    \resizebox{0.70\textwidth}{!}{
	\begin{tabular}{llccc|cc|cc|cc}
        \hline
        \hline
        \vspace{0.08cm}
		Name in this work & Name in catalog & Region & $N_{\mathrm{c}}$ & Age$_{\mathrm{Chronos}}$ & $M^{\mathrm{cls}}$ & $M^{\mathrm{cls}}_*$ & $M^{\mathrm{cloud}}_{\mathrm{SFE=1\%}}$ &  $R_{\mathrm{SFE=1\%}}$ & $M^{\mathrm{cloud}}_{\mathrm{SFE=3\%}}$ & $R_{\mathrm{SFE=3\%}}$ \\
  		 &  &  & & $\mathrm{[Myr]}$ & $\mathrm{[M_\odot]}$ & $\mathrm{[M_\odot]}$ & $\mathrm{[M_\odot]}$ & $\mathrm{[pc]}$  & $\mathrm{[M_\odot]}$ & $\mathrm{[pc]}$  \\
        \hline 
        Briceno\,1 & Briceno\,1 & Orion & 171 & $12.7_{-0.1}^{+1.5}$ & 107 & 205 & 20500 & 15.7 & 6833 & 10.7 \\
        OBP-West & ASCC\,18 & Orion & 82 & $12.8_{-1.7}^{+1.4}$ & 64 & 155 & 15500 & 14.2 & 5167 & 9.7 \\
        OBP-d & Theia\,13 & Orion & 249 & $10.3_{-1.1}^{+1.0}$ & 156 & 340 & 34000 & 18.7 & 11333 & 12.8 \\
        Sigma\,Orionis & Sigma\,Orionis & Orion & 181 & $3.4_{-0.5}^{+0.7}$ & 120 & 188 & 18750 & 15.2 & 6250 & 10.4 \\
        NGC\,1980 & NGC\,1980 & Orion & 364 & $8.1_{-0.8}^{+0.4}$ & 226 & 490 & 49000 & 21.2 & 16333 & 14.5 \\
        NGC\,1981 & UBC\,207 & Orion & 53 & $6.0_{-0.7}^{+1.5}$ & 34 & 92 & 9250 & 11.9 & 3083 & 8.2 \\
        NGC\,1977 & NGC\,1977 & Orion & 111 & $6.8_{-1.1}^{+1.6}$ & 92 & 255 & 25500 & 16.9 & 8500 & 11.6 \\
        \hline
        OBP-Near-1 & ASCC\,19 & Orion & 61 & $12.5_{-1.5}^{+0.4}$ & 51 & 115 & 11500 & 12.8 & 3833 & 8.8 \\
        ASCC\,20 & ASCC\,20 & Orion & 194 & $19.3_{-0.2}^{+6.6}$ & 138 & 255 & 25500 & 16.9 & 8500 & 11.6 \\
        ASCC\,21 & ASCC\,21 & Orion & 116 & $12.6_{-1.8}^{+0.2}$ & 102 & 198 & 19750 & 15.5 & 6583 & 10.6 \\
        Heleus & Alessi-Teutsch\,10 & Perseus & 85 & $6.4_{-0.8}^{+1.1}$ & 52 & 82 & 8250 & 11.5 & 2750 & 7.8 \\
        IC2118-Halo & CWNU\,1028 & Orion & 19 & $13.4_{-3.2}^{+3.7}$ & 15 & 30 & 3000 & 8.1 & 1000 & 5.5 \\
        Orion-A-East & CWNU\,1072 & Orion & 60 & $20.5_{-3.3}^{+1.2}$ & 34 & 85 & 8500 & 11.6 & 2833 & 7.9 \\
        L1630-background & CWNU\,1088 & Orion & 46 & $1.7_{-0.1}^{+1.4}$ & 17 & 28 & 2750 & 7.8 & 917 & 5.4 \\
        L1598-East & CWNU\,1092 & Orion & 27 & $25.0_{-0.6}^{+18.2}$ & 24 & 48 & 4750 & 9.5 & 1583 & 6.5 \\
        L1598 & CWNU\,1106 & Orion & 16 & $5.5_{-0.9}^{+0.8}$ & 10 & 25 & 2500 & 7.6 & 833 & 5.2 \\
        L1546 & CWNU\,1129 & Taurus & 34 & $1.9_{-0.3}^{+1.2}$ & 7 & 8 & 750 & 5.0 & 250 & 3.4 \\
        lambda-Ori & Collinder\,69 & Orion & 1247 & $7.2_{-0.2}^{+0.7}$ & 741 & 1442 & 144250 & 30.7 & 48083 & 21.0 \\
        Autochthe-Gorgophone & HSC\,1250 & Perseus & 34 & $6.8_{-1.4}^{+3.9}$ & 27 & 55 & 5500 & 10.0 & 1833 & 6.8 \\
        Mestor & HSC\,1262 & Perseus & 143 & $8.4_{-2.2}^{+0.7}$ & 81 & 150 & 15000 & 14.1 & 5000 & 9.6 \\
        L1495 & HSC\,1318 & Taurus & 52 & $4.7_{-1.5}^{+1.6}$ & 24 & 40 & 4000 & 8.9 & 1333 & 6.1 \\
        HSC\,1340 & HSC\,1340 & Taurus & 194 & $27.2_{-2.3}^{+4.1}$ & 98 & 220 & 22000 & 16.1 & 7333 & 11.0 \\
        HSC\,1373 & HSC\,1373 & Taurus & 46 & $21.3_{-2.5}^{+8.9}$ & 20 & 40 & 4000 & 8.9 & 1333 & 6.1 \\
        HSC\,1481 & HSC\,1481 & Taurus & 40 & $22.4_{-3.4}^{+15.5}$ & 15 & 28 & 2750 & 7.8 & 917 & 5.4 \\
        L1634-North & HSC\,1633 & Orion & 68 & $9.2_{-2.9}^{+0.6}$ & 42 & 98 & 9750 & 12.1 & 3250 & 8.3 \\
        Eridanus-North & HSC\,1640 & Orion & 128 & $15.5_{-0.7}^{+2.6}$ & 64 & 145 & 14500 & 13.9 & 4833 & 9.5 \\
        Rigel & HSC\,1648 & Orion & 78 & $13.0_{-1.6}^{+0.8}$ & 39 & 65 & 6500 & 10.6 & 2167 & 7.2 \\
        L1634-South & HSC\,1653 & Orion & 30 & $10.0_{-1.6}^{+9.6}$ & 16 & 32 & 3250 & 8.3 & 1083 & 5.7 \\
        HSC\,1692 & HSC\,1692 & Orion & 24 & $27.7_{-1.9}^{+5.4}$ & 13 & 30 & 3000 & 8.1 & 1000 & 5.5 \\
        IC\,348 & IC\,348 & Perseus & 302 & $4.9_{-2.2}^{+1.2}$ & 151 & 295 & 29500 & 17.8 & 9833 & 12.2 \\
        L1641-South & L\,1641S & Orion & 72 & $9.0_{-2.6}^{+2.0}$ & 53 & 155 & 15500 & 14.2 & 5167 & 9.7 \\
        Mamajek\,3 & Mamajek\,3 & Taurus & 33 & $17.6_{-0.8}^{+9.6}$ & 19 & 42 & 4250 & 9.1 & 1417 & 6.2 \\
        NGC\,1333 & NGC\,1333 & Perseus & 31 & $3.8_{-0.9}^{+1.9}$ & 10 & 10 & 1000 & 5.5 & 333 & 3.8 \\
        NGC\,2068 & NGC\,2068 & Orion & 102 & $3.3_{-0.8}^{+0.7}$ & 65 & 142 & 14250 & 13.8 & 4750 & 9.5 \\
        OC\,0340 & OC\,0340 & Orion & 17 & $13.4_{-4.9}^{+3.3}$ & 30 & 20 & 2000 & 7.0 & 667 & 4.8 \\
        OCSN\,50 & OCSN\,50 & Taurus & 24 & $18.1_{-5.2}^{+8.1}$ & 13 & 30 & 3000 & 8.1 & 1000 & 5.5 \\
        omega-Ori & OCSN\,56 & Orion & 88 & $17.7_{-1.8}^{+2.5}$ & 57 & 130 & 13000 & 13.4 & 4333 & 9.2 \\
        L1616 & OCSN\,59 & Orion & 60 & $8.3_{-0.9}^{+2.0}$ & 36 & 75 & 7500 & 11.1 & 2500 & 7.6 \\
        OBP-b & OCSN\,61 & Orion & 147 & $19.6_{-1.4}^{+0.3}$ & 94 & 170 & 17000 & 14.7 & 5667 & 10.1 \\
        OBP-e & OCSN\,64 & Orion & 67 & $18.2_{-2.3}^{+6.9}$ & 45 & 82 & 8250 & 11.5 & 2750 & 7.8 \\
        OBP-far & OCSN\,65 & Orion & 70 & $17.1_{-2.4}^{+3.4}$ & 43 & 95 & 9500 & 12.0 & 3167 & 8.2 \\
        OCSN\,68 & OCSN\,68 & Orion & 51 & $20.5_{-5.5}^{+1.7}$ & 29 & 68 & 6750 & 10.7 & 2250 & 7.3 \\
        L1647-North & OCSN\,70 & Orion & 18 & $9.9_{-6.4}^{+7.2}$ & 15 & 38 & 3750 & 8.7 & 1250 & 6.0 \\
        Alcaeus & OC\,0279 & Perseus & 146 & $13.4_{-1.3}^{+5.6}$ & 82 & 145 & 14500 & 13.9 & 4833 & 9.5 \\
        Electryon-Cynurus & OC\,0280 & Perseus & 78 & $13.5_{-5.1}^{+3.0}$ & 58 & 152 & 15250 & 14.2 & 5083 & 9.7 \\
        OBP-Near-3 & OC\,0339 & Orion & 19 & $12.9_{-5.5}^{+11.0}$ & 10 & 20 & 2000 & 7.0 & 667 & 4.8 \\
        L1647-Main & OC\,0356 & Orion & 15 & $5.7_{-2.2}^{+2.3}$ & 9 & 25 & 2500 & 7.6 & 833 & 5.2 \\
        L1517 & Theia\,54 & Taurus & 48 & $7.7_{-2.8}^{+2.0}$ & 33 & 52 & 5250 & 9.8 & 1750 & 6.7 \\
        118Tau & Theia\,65 & Taurus & 37 & $14.1_{-2.1}^{+2.7}$ & 17 & 32 & 3250 & 8.3 & 1083 & 5.7 \\
        L1551 & Theia\,66 & Taurus & 39 & $5.1_{-1.3}^{+2.4}$ & 19 & 30 & 3000 & 8.1 & 1000 & 5.5 \\
        L1524 & Theia\,7 & Taurus & 47 & $8.5_{-3.4}^{+2.7}$ & 26 & 65 & 6500 & 10.6 & 2167 & 7.2 \\
        L1544 & Theia\,93 & Taurus & 91 & $10.7_{-1.9}^{+2.0}$ & 49 & 128 & 12750 & 13.3 & 4250 & 9.1 \\
        OBP-Near-2 & UBC\,17a & Orion & 102 & $6.3_{-0.5}^{+0.5}$ & 70 & 118 & 11750 & 12.9 & 3917 & 8.9 \\
        lambda-Ori-South & UPK\,398 & Orion & 84 & $10.1_{-1.1}^{+4.1}$ & 45 & 95 & 9500 & 12.0 & 3167 & 8.2 \\
        L1617 & UPK\,402 & Orion & 47 & $5.9_{-1.8}^{+1.3}$ & 27 & 62 & 6250 & 10.4 & 2083 & 7.1 \\
        Orion-Y & UPK\,422 & Orion & 240 & $25.1_{-6.3}^{+0.6}$ & 140 & 300 & 30000 & 17.9 & 10000 & 12.2 \\
		\hline
	\end{tabular}}
    \tablefoot{For each cluster, the following information is provided: the name used in this work and the one from the source catalogs \citep{{Hunt2023,Cantat-Gaudin2020,Sim2019,Hao2022,He2022,Szilagyi2021,Liu2019}}, the region to which they are associated, the number of stellar members ($N_{\mathrm{c}}$), the estimated isochronal age (Age$_{\mathrm{Chronos}}$), the cluster's mass derived from the stellar members ($M^{\mathrm{cls}}$), and corrected for incompleteness ($M^{\mathrm{cls}}_*$). Additionally, the mass and radius of the cloud associated with the cluster are given assuming a SFE of 1\% and 3\% ($M^{\mathrm{cloud}}_{\mathrm{SFE=1(3)\%}}$, $R_{\mathrm{SFE=1(3)\%}}$).}
	\label{tab:RWclusters-properties}
\end{table*}

\correction{In this appendix, we provide supplementary plots and a comparison with the literature for the ages and masses of the Radcliffe wave's clusters used in this study. In Table~\ref{tab:RWclusters-6Dinfo-allClusters}, the initial positions and velocities of these clusters, along with the errors, are listed. Additionally, we describe our method for estimating the number of past supernova (SN) events.}

\subsection{Age computation}\label{app:age_computation}

\correction{In Sect.~\ref{sec:age_mass_computation}, we outline the method used to compute the ages of the clusters.}
The color-magnitude diagrams for the 56 clusters of the Radcliffe wave used in this study are shown in Fig.~\href{https://doi.org/10.5281/zenodo.14626660}{A.1} and \href{https://doi.org/10.5281/zenodo.14626660}{A.2}. The estimated ages are listed in Table~\ref{tab:RWclusters-properties}.

In Fig.~\href{https://doi.org/10.5281/zenodo.14626660}{A.3}, we present the comparison between the ages of the clusters computed with {\tt Chronos} \correction{\citep{Ratzenboech2023b}} and those provided by the source catalogs. 
It is possible to note that, in general, the estimated ages are consistent with each other. Exceptions are observed for certain clusters, namely 
CWNU 1088 ($\mathrm{age}_{\mathrm{Chronos}} = 1.7_{-0.1}^{+1.4}\,\mathrm{Myr}$; $\mathrm{age}_{\mathrm{catalog}} = 30_{-19}^{+34}\,\mathrm{Myr}$), 
L1524 ($\mathrm{age}_{\mathrm{Chronos}} = 8.5_{-3.4}^{+2.7}\,\mathrm{Myr}$; $\mathrm{age}_{\mathrm{catalog}} = 187_{-137}^{+808}\,\mathrm{Myr}$), 
L1546 ($\mathrm{age}_{\mathrm{Chronos}} = 1.9_{-0.3}^{+1.1}\,\mathrm{Myr}$; $\mathrm{age}_{\mathrm{catalog}} =15_{-11}^{+22}\,\mathrm{Myr}$), 
and NGC 1977 ($\mathrm{age}_{\mathrm{Chronos}} = 6.8_{-1.1}^{+1.6}\,\mathrm{Myr}$; $\mathrm{age}_{\mathrm{catalog}} =97.7\,\mathrm{Myr}$–error not provided) 
for which the ages calculated with {\tt Chronos} are notably younger. In addition, given the estimated age uncertainty, the {\tt Chronos}’ ages are likely more precise when compared to the ones given in the cluster catalog. Moreover, some of the selected clusters have already been studied in \correction{other} literature, and their computed ages are consistent with our results. For instance, the ages of OBP-West (also known as ASCC 18), ASCC 20, and ASCC 21 have been estimated \citep{Kos2019} to be about $12.75 \pm 1.27\,\mathrm{Myr}$, $21.25 \pm 2.12\,\mathrm{Myr}$, and $11.0 \pm 1.1\,\mathrm{Myr}$  respectively. In our analysis, these clusters are  $12.8_{-1.7}^{+1.4}$, $19.3_{-0.2}^{+6.6}$, and $12.6_{-1.8}^{+0.2}\,\mathrm{Myr}$ old. The age of NGC 1980 has been computed \citep{Alves2012} to be between 5 and 10 Myr, in agreement with our range of 7.3 and 8.5 Myr.

\subsection{Mass computation}\label{app:mass_computation}

\correction{The masses of the clusters are estimated as described in Sect.~\ref{sec:age_mass_computation} and reported in Table~\ref{tab:RWclusters-properties}.}
In Fig.~\href{https://doi.org/10.5281/zenodo.14626660}{A.4}, we show the best-fit mass functions, together with the observed ones, for each of the analyzed clusters. \correction{The plots show that all the populations} exhibit a truncation at the low-mass end caused by sensitivity limits. We also observe that the mass distributions are adequately sampled up to several Solar masses for some clusters, whereas for others, they are truncated (e.g., HSC 1640). This truncation may be attributed to catalog incompleteness \correction{(too bright sources)} or stellar evolution processes \correction{(SN in the past)}. 

By comparing the masses of our clusters with those found in the literature \correction{\citep[see e.g.,][]{Alves2012,Almeida2023}}, we confirm that our values are generally lower estimates, as \correction{HR23} prioritized precision over completeness, aiming to minimize the number of false positives associated with each cluster. For example, the mass of Briceno 1 (also known as ASCC 16) has been estimated \correction{to have about} $441 \pm 88 \, \mathrm{M}_{\odot}$ \citep{Almeida2023}, more than double our own estimation of $205 \, \mathrm{M}_{\odot}$. The same happens for NGC 1980, for which a mass of $1000 \, \mathrm{M}_{\odot}$ has been computed \citep{Alves2012}, \correction{twice as much than} what we obtain. Instead, for OBP-d (also known as Theia-13) the estimated mass of $365 \pm 73 \, \mathrm{M}_{\odot}$ \citep{Almeida2023}, aligns closely with our estimate of $340 \, \mathrm{M}_{\odot}$.

\subsection{Estimation of the past supernova events}\label{app:SN_computation}

\correction{To determine whether radionuclides were present during the encounter of the Sun with the Radcliffe wave, we estimate the number of possible SN that occurred in the region of interest before $11.5 \pm 0.3$ Myr ago. This time threshold is chosen because it corresponds to the estimated period when the crossing of the Radcliffe wave likely ended (assuming cloud radii of 50 pc).}
We focused on 37 out of the 56 clusters, which are the ones that are currently located in the Orion star-forming region, as this is the region to which the Sun gets closer. 

For a given cluster, we used the estimated mass corrected for incompleteness (see Sect.~\ref{sec:age_mass_computation}) and Kroupa IMF \citep{Kroupa2001}, to generate its stellar content. We then counted the number of stars with a mass greater than the highest one predicted by a stellar isochrone model \citep{Bressan2012,Nguyen2022} corresponding to an age equal to that of the analyzed cluster 11.5 Myr ago. Clusters that are younger than 11.5 Myr were not considered. \correction{In this way, we were able to roughly estimate the number of massive stars that had time to evolve and, eventually, explode as a SN.}
We repeated this process multiple times in order to account for errors in the clusters' ages, the timing of the Sun exiting the Radcliffe wave, and the variability in the number of stars produced by the IMF. We estimate that approximately $3_{-2}^{+1}$ SN occurred before 11.5 Myr ago, suggesting that the Sun passed through freshly enriched clouds. This value, in line with studies on SN in young clusters \citep{Foley2023}, is likely underestimated due to catalog incompleteness, which consequently affects the mass estimates of the clusters. If we hypothesize that half of the actual clusters mass is missing from our estimates, as found in some cases, the number of SN would be $8_{-4}^{+1}$.

\section{Initial condition tests}\label{app:initialCondTest}

\correction{In this appendix, we present how different initial conditions influence our results. In particular, we used different solar parameters and various Milky Way potentials from the literature.}
We performed multiple sets of orbit integrations using the procedure described in Sect.~\ref{sec:orbits_traceback}, varying a parameter or a combination of parameters in each run, to evaluate how the crossing times of the Solar System with the Radcliffe wave are influenced by them.

\correction{In our main study, we used {\tt galpy}’s {\tt MWPotential2014} as a model for the Milky Way potential. This model includes a bulge, a disk, and a halo component \citep[][]{Bovy2015}. We assumed the distance of the Sun from the Galactic center to be $R_{\odot} = 8.33 \, \mathrm{kpc}$ \citep{Gillessen2009}, its vertical position respect to the disk to be  $z_{\odot}= 27 \, \mathrm{pc}$ \citep{Chen2001}, and its velocity relative to the Local Standard of Rest ($v_{\mathrm{LSR}} = 220 \, \mathrm{km}\,\mathrm{s}^{-1}$) to be ($U_\odot,\, V_\odot,\, W_\odot$) = (11.1, 12.24, 7.25)$\,\mathrm{km}\,\mathrm{s}^{-1}$ \citep{Schoenrich2010}.}

In an alternative set-up, we examined the scenario where the Sun’s Galactocentric distance and height above the plane are set to $R_\odot = 8.122 \, \mathrm{kpc}$ \citep{GRAVITY2018} and $z_\odot= 20.8 \, \mathrm{pc}$ \citep{Bennett2019}, respectively. Moreover, we explored two other models of the Milky Way’s potential included in {\tt galpy}. In one model \citep{McMillan2017}, the Local Standard of Rest has a circular velocity of $233\,\mathrm{km}\,\mathrm{s}^{-1}$ and the Galactocentric radius is $R_\odot = 8.2 \, \mathrm{kpc}$; while in the second model \citep{Irrgang2013}, the Local Standard of Rest has a circular velocity of $242\,\mathrm{km}\,\mathrm{s}^{-1}$, with $R_\odot = 8.4 \, \mathrm{kpc}$. Moreover, we investigated the effect that different peculiar velocities of the Sun would have on our results, using ($U_\odot,\, V_\odot,\, W_\odot$) = (10.1, 4.0, 6.7)$\,\mathrm{km}\,\mathrm{s}^{-1}$ \citep{Hogg2005}, (10.0, 5.25, 7.17)$\,\mathrm{km}\,\mathrm{s}^{-1}$ \citep{Dehnen1998}, or (10.0, 15.4, 7.8)$\,\mathrm{km}\,\mathrm{s}^{-1}$ \citep{Kerr1986}.
The full list of cases analyzed, along with the computed transit times between the Sun and the Radcliffe wave for various threshold distances ($d_{\mathrm{Sun-cloud}}$), is provided in Table~\ref{tab:initial_condition_test}.
From these tests, we conclude that over the 30 Myr integration period, our results are robust and do not vary significantly when considering different initial conditions.

\begin{sidewaystable*}
    \scriptsize
	\centering
	\caption{Test of initial conditions to assess the robustness of our findings.}
	\label{tab:initial_condition_test}
    \renewcommand{\arraystretch}{1.2}
	\begin{tabular}{cccccc|cccc} 
        \hline
        \hline
        \vspace{0.08cm}
		\correction{Parameter Investigated} & Potential & $d_{\odot}$ & $z_\odot$ & ($U_{\odot},\,V_{\odot},\,W_{\odot}$) & Ref.   & $d_{\mathrm{Sun-cloud}}\le50\,\mathrm{pc}$ & $d_{\mathrm{Sun-cloud}}\le40\,\mathrm{pc}$ & $d_{\mathrm{Sun-cloud}}\le30\,\mathrm{pc}$ & $d_{\mathrm{Sun-cloud}}\le20\,\mathrm{pc}$ \\
              &           & [kpc]       & [pc]      & [$\mathrm{km\,s^{-1}}$]               &        & ($t_{\mathrm{enter}}$, $t_{\mathrm{exit}}$) [Myr] & ($t_{\mathrm{enter}}$, $t_{\mathrm{exit}}$) [Myr] & ($t_{\mathrm{enter}}$, $t_{\mathrm{exit}}$) [Myr] & ($t_{\mathrm{enter}}$, $t_{\mathrm{exit}}$) [Myr] \\  
		\hline 
        Ref.Case & mwp14$^{*}$ & 8.3 & 27 & (11.1,\,12.24,\,7.25) & a$_1$,b$_1$,c$_1$,d$_1$ & 
        ($-18.2\pm0.1$, $-11.5\pm0.3$) & ($-17.4\pm0.2$, $-11.9\pm0.3$) & ($-14.8\pm0.1$, $-12.4\pm0.2$) & ($-14.3\pm0.1$, $-12.8\pm0.2$) \\
        \hline
        {Sun Galactocentric dist.}
         & mwp14 & 7.5   & 27 & (11.1,\,12.24,\,7.25)  & a$_1$,b$_2$,c$_1$,d$_1$ & 
        ($-18.0\pm0.1$, $-11.5\pm0.6$) & ($-15.2\pm0.2$, $-12.1\pm0.3$) & ($-14.6\pm0.2$, $-12.6\pm0.2$) & ($-14.5\pm0.2$, $-12.6\pm0.3$)$_{d\le27.5\,\mathrm{pc}}$ \\
         & mwp14 & 7.94  & 27 & (11.1,\,12.24,\,7.25)  & a$_1$,b$_3$,c$_1$,d$_1$ &
        ($-18.2\pm0.1$, $-11.5\pm0.6$) & ($-17.3\pm0.2$, $-12.0\pm0.3$) & ($-14.8\pm0.1$, $-12.4\pm0.2$) & ($-14.4\pm0.1$, $-12.7\pm0.3$)$_{d\le22.5\,\mathrm{pc}}$ \\
         & mwp14 & 8.122 & 27 & (11.1,\,12.24,\,7.25)  & a$_1$,b$_4$,c$_1$,d$_1$ &
        ($-18.2\pm0.1$, $-11.4\pm0.3$) & ($-17.4\pm0.2$, $-11.8\pm0.3$) & ($-14.8\pm0.1$, $-12.4\pm0.2$) & ($-14.2\pm0.1$, $-12.8\pm0.2$) \\
         & mwp14 & 8.7   & 27 & (11.1,\,12.24,\,7.25)  & a$_1$,b$_5$,c$_1$,d$_1$ &
        ($-18.3\pm0.2$, $-11.4\pm0.3$) & ($-17.5\pm0.3$, $-11.4\pm0.5$) & ($-14.9\pm0.1$, $-12.3\pm0.2$) & ($-14.4\pm0.1$, $-12.7\pm0.1$) \\
        \hline
        {Sun velocity}
         & mwp14 & 8.3 & 27 & (10.1,\,4.0,\,6.7)    & a$_1$,b$_1$,c$_1$,d$_2$ &
        ($-18.3\pm0.1$, $-11.4\pm0.3$) & ($-17.4\pm0.2$, $-11.9\pm0.3$) & ($-14.8\pm0.1$, $-12.4\pm0.2$) & ($-14.3\pm0.1$, $-12.8\pm0.2$) \\
         & mwp14 & 8.3 & 27 & (10.0,\,5.25,\,7.17)  & a$_1$,b$_1$,c$_1$,d$_3$ &
        ($-18.3\pm0.1$, $-11.4\pm0.3$) & ($-17.4\pm0.2$, $-11.9\pm0.3$) & ($-14.7\pm0.1$, $-12.4\pm0.2$) & ($-14.3\pm0.1$, $-12.8\pm0.2$) \\
         & mwp14 & 8.3 & 27 & (10.0,\,15.4,\,7.8)   & a$_1$,b$_1$,c$_1$,d$_4$ &
        ($-18.3\pm0.1$, $-11.4\pm0.2$) & ($-17.4\pm0.3$, $-11.9\pm0.3$) & ($-14.9\pm0.1$, $-12.4\pm0.2$) & ($-14.3\pm0.1$, $-12.8\pm0.2$) \\
        \hline
        {Sun height}
         & mwp14 & 8.3 & 0  & (11.1,\,12.24,\,7.25) & a$_1$,b$_1$,-,d$_1$ &
        ($-18.4\pm0.2$, $-11.4\pm0.2$) & ($-17.6\pm0.2$, $-11.4\pm0.8$) & ($-14.9\pm0.1$, $-12.4\pm0.1$) & ($-14.4\pm0.1$, $-12.8\pm0.2$) \\
         & mwp14 & 8.3 & 50 & (11.1,\,12.24,\,7.25) & a$_1$,b$_1$,-,d$_1$ &
        ($-18.2\pm0.1$, $-11.5\pm0.3$) & ($-17.4\pm0.2$, $-11.9\pm0.3$) & ($-14.8\pm0.1$, $-12.4\pm0.2$) & ($-14.3\pm0.1$, $-12.8\pm0.2$) \\
        \hline
        {MW's potential}
         & McMillan17$^{**}$ & 8.2  & 20.8 & (11.1,\,12.24,\,7.25) & a$_2$,a$_2$,c$_2$,d$_1$ &
         ($-18.2\pm0.1$, $-11.4\pm0.2$) & ($-17.3\pm0.2$, $-11.9\pm0.3$) & ($-14.8\pm0.1$, $-12.3\pm0.2$) & ($-14.1\pm0.1$, $-12.8\pm0.2$)\\
         & Irrgang13I$^{***}$ & 8.4 & 20.8 & (11.1,\,12.24,\,7.25) & a$_3$,a$_3$,c$_2$,d$_1$ &
        ($-18.2\pm0.1$, $-11.5\pm0.3$) & ($-17.3\pm0.2$, $-11.8\pm0.3$) & ($-14.8\pm0.1$, $-12.4\pm0.2$) & ($-14.3\pm0.1$, $-12.8\pm0.3$) \\
        \hline
        {Mixed}
         & mwp14 & 8.122 & 20.8 & (11.1,\,12.24,\,7.25) & a$_1$,b$_4$,c$_2$,d$_1$ &
        ($-18.3\pm0.1$, $-11.5\pm0.3$) & ($-17.4\pm0.2$, $-12.0\pm0.3$) & ($-14.8\pm0.1$, $-12.4\pm0.2$) & ($-14.2\pm0.1$, $-12.9\pm0.2$)\\
         & mwp14 & 8.122 & 20.8 & (10.0,\,15.4,\,7.8)   & a$_1$,b$_4$,c$_2$,d$_4$ &
        ($-18.3\pm0.1$, $-11.5\pm0.3$) & ($-17.4\pm0.3$, $-12.0\pm0.1$) & ($-14.9\pm0.1$, $-12.4\pm0.2$) & ($-14.2\pm0.1$, $-12.9\pm0.2$)\\
		\hline
	\end{tabular}
    \tablefoot{The time intervals for the encounters between the Solar System and the Radcliffe wave were calculated by varying the initial parameters of orbit integration. The investigated initial conditions include the Sun's Galactocentric distance, the Sun's velocity relative to the Local Standard of Rest, the height above the disk midplane of the Sun, and the potential of the Milky Way. The \correction{statistical} errors of the time intervals are computed through Monte Carlo sampling the uncertainty distributions of the initial positions and velocities of the clusters, as well as the assumed initial parameters of the Sun. The errors for the Sun's initial parameters are \correction{taken from the given references.} In cases where no encounter occurs for a given distance, the reported time range corresponds to the distance indicated as a subscript. In the Reference column, the letters $a$, $b$, $c$, and $d$ refer to the Milky Way potential, Galactocentric distance, Sun height above the disk, and Sun's velocity relative to the Local Standard of Rest, respectively.\\
    The circular velocity at the Sun's position is set to:
                            $^{*}$   $v_{\mathrm{LSR}}\,=\,220\,\mathrm{km\,s^{-1}}$; 
                            $^{**}$  $v_{\mathrm{LSR}}\,=\,233.1\,\mathrm{km\,s^{-1}}$;
                            $^{***}$ $v_{\mathrm{LSR}}\,=\,242\,\mathrm{km\,s^{-1}}$.}
    \tablebib{
    (a$_1$) \cite{Bovy2015}; (a$_2$) \cite{McMillan2017}; (a$_3$) \cite{Irrgang2013};
    (b$_1$) \cite{Gillessen2009}; (b$_2$) \cite{Francis2014}; (b$_3$) \cite{Eisenhauer2003}; (b$_4$) \cite{GRAVITY2018}; (b$_5$) \cite{Vanhollebeke2009}; 
    (c$_1$) \cite{Chen2001}; (c$_2$) \cite{Bennett2019};
    (d$_1$) \cite{Schoenrich2010}; (d$_2$) \cite{Hogg2005}; (d$_3$) \cite{Dehnen1998}; (d$_4$) \cite{Kerr1986}.
    }
\end{sidewaystable*}

\end{appendix}

\end{document}